\pdfoutput=1
\documentclass[american,aps,pra,reprint,superscriptaddress,twocolumn]{revtex4-1}
\usepackage{amsmath,bbm}
\usepackage{amsthm}
\usepackage{amsmath}
\usepackage{latexsym}
\usepackage{amssymb}
\usepackage{float}
\usepackage{graphicx}           
\usepackage{color}
\usepackage{mathpazo}
\usepackage{comment}
\usepackage{mathtools}

\usepackage[colorlinks=true,linkcolor=blue,citecolor=magenta,urlcolor=blue]{hyperref}

\newcommand{\be}{\begin{equation}}
\newcommand{\ee}{\end{equation}}
\newcommand{\bea}{\begin{eqnarray}}
\newcommand{\eea}{\end{eqnarray}}
\allowdisplaybreaks

\def\squareforqed{\hbox{\rlap{$\sqcap$}$\sqcup$}}
\def\qed{\ifmmode\squareforqed\else{\unskip\nobreak\hfil
\penalty50\hskip1em\null\nobreak\hfil\squareforqed
\parfillskip=0pt\finalhyphendemerits=0\endgraf}\fi}
\def\endenv{\ifmmode\;\else{\unskip\nobreak\hfil
\penalty50\hskip1em\null\nobreak\hfil\;
\parfillskip=0pt\finalhyphendemerits=0\endgraf}\fi}

\makeatletter
\newtheorem*{rep@theorem}{\rep@title}
\newcommand{\newreptheorem}[2]{%
\newenvironment{rep#1}[1]{%
 \def\rep@title{#2 \ref{##1}}%
 \begin{rep@theorem}}%
 {\end{rep@theorem}}}
\makeatother

\newreptheorem{thm}{Theorem}

\begin{document}

\title{Role of nonclassical temporal correlation in powering quantum random access codes}

\author{Subhankar Bera}
\email{berasanu007@gmail.com}
\affiliation{S. N. Bose National Centre for Basic Sciences, Block JD, Sector III, Salt Lake, Kolkata 700106, India }

\author{Ananda G. Maity}
\email{anandamaity289@gmail.com}
\affiliation{S. N. Bose National Centre for Basic Sciences, Block JD, Sector III, Salt Lake, Kolkata 700106, India }

\author{Shiladitya Mal}
\email{shiladitya.27@gmail.com}
\affiliation{Physics Division, National Center for Theoretical Sciences, Taipei 10617, Taiwan}
\affiliation{Department of Physics and Center for Quantum Frontiers of Research and Technology (QFort), National Cheng Kung University, Tainan 701, Taiwan}

\author{A. S. Majumdar}
\email{archan@bose.res.in}
\affiliation{S. N. Bose National Centre for Basic Sciences, Block JD, Sector III, Salt Lake, Kolkata 700106, India }

\begin{abstract}
We explore the fundamental origin of the quantum advantage behind random access code. We propose new temporal inequalities compatible with noninvasive-realist models and show that any non-zero quantum advantage of $n \mapsto 1$ random access code in presence of shared randomness is equivalent to the violation of the corresponding temporal inequality. As a consequence of this connection we also prove that maximal success probability of $n \mapsto 1$ random access code can be obtained when the maximal violation of the corresponding inequality is achieved. We further show that any non-zero quantum advantage of $n \mapsto 1$ random access code, or in other words, any non-zero violation of the corresponding temporal inequality can certify genuine randomness.
\end{abstract}

\maketitle

\section{Introduction} 
Random access codes (RAC) are among the most fundamental and powerful communication scenarios where a sender encodes a  string of message into fewer bits and  sends it to the receiver who then aims to recover any of the initial bits with some good probability of success \cite{raco1,raco2,raco3,Nayak99,Ambainis_SR}. In a general $n \mapsto m$ RAC, there are $n$ number of bits (say, $x_1,x_2,...,x_n$) of message which the sender (say, Alice)  sends to the receiver (say, Bob) by encoding the message in a $m$ bit string (where $m<n$). In each round, Bob picks out a random number $i$ (where $i \in \lbrace 1,2,...,n \rbrace$) and tries to guess the $i$-th bit ($x_i$) of Alice. The probability with which Bob can decode the message can be increased if instead of a classical system, Alice sends a quantum system, e.g., by using either quantum communication \cite{raco3}, or communication of classical bits assisted with a shared quantum state\cite{erac,srac1,srac2,Das21}. Such quantum random access codes have been introduced and developed for qubit systems as well as higher dimensional quantum systems.

Historically, RAC was first proposed in order to show the enormous information carrying capabilities of a quantum system compared to a classical system of 
the same dimension. Although the well-known Holevo bound \cite{holevo} 
stipulates that by using a qubit system one cannot transmit more information reliably to the receiver than one classical bit, this does not solely 
portray the information carrying capabilities of a quantum system. A $m$-qubit can in general be depicted by an unit vector in $2^m$ dimensional complex Hilbert space revealing the possibility of encoding classical information with exponentially fewer qubits. In general, Bob may not need to know the information of all $n$ bits together, but rather may choose to extract some bits of classical information out of the encoding depending on some task and therefore, may explore some degrees of freedom which otherwise were frozen. This motivates to formulate the task of RAC without contradicting Holevo's result.  Random access codes have been shown to possess wide range of applications such as quantum finite automata \cite{raco2,raco3,Nayak99}, network coding \cite{netcod1,netcod}, locally decodable codes \cite{ldc1,ldc2,ldc3}, non-local games\cite{Bridge}, dimension witness \cite{dw,dw2,dw3,dw4}, quantum communication complexity \cite{ccx,Aaronson04,Gavinsky06,Buhrman01,Mar18}, randomness certification \cite{rangen}, quantum cryptography \cite{qkd}, studies of no-signaling resources \cite{Grudka}, self-testing of quantum measurements \cite{Mohan}, and so on. 

However, there are also some surprising results such that although $n\mapsto 1$ 
quantum RAC with shared randomness (SR) exists \cite{Ambainis_SR}, but it does not exist without SR for $n \geq 4$ \cite{netcod1}. A variant of this communication task was introduced by Spekkens {\it et al} \cite{exprac1} under a cryptographic constraint where preparation contextuality is shown to be the actual cause for achieving the quantum advantages (see also \cite{alok1}). Recently an analytical way of finding maximal quantum bound for preparation non-contextuality inequality has been derived \cite{Sharma}.  Investigating the fundamental cause behind the quantum advantage of RAC is not only important from the foundational perspective, but may also escalate the way of finding new applications in information processing and communication tasks.

It is well known from the celebrated Bell theorem \cite{Bell'64,chsh'69} that correlations between spatially separated events are more restricted in classical theory than allowed in quantum theory. On the other hand, two fundamental no go theorems for time-like separated events  involve non-contextual hidden variable models (NCVM) \cite{Bell'66,Kochen'67} and macro-realism \cite{LG'85}, compatible with classical theory. macro-realism (MR) was introduced by Leggett and Garg to probe quantumness of `macroscopic' systems. MR is a conjunction of two assumptions, {\it macro-realism per se} and {\it noninvasive measurability} \cite{LG'85,LG'02,emary'14,yearsley}. A  variant of MR is known as non-invasive realist model where {\it macro-realism per se} is replaced by {\it realism}. %which asserts that `at any instant a system is definitely in any one of the available distinguishable states such that all its observable properties have definite values'. On the other hand, {\it Noninvasive measurability} implies that `it is possible, in principle, to determine which of the states the system is in, without affecting the state itself or the system's subsequent evolution'. 
Quantum theory violates consequences of both the models, i.e., KCBS inequality derivable from NCVM \cite{KCBS} and Legget-Garg inequality (LGI) arising from MR \cite{LG'85}. In recent years, LGI and  other temporal correlations have  acquired considerable attention from both fundamental \cite{yearsley,kofler'07,Naik'22,kofler'13,brukner'04,mal'16,mal'18,das'16,mal'19,fritz'10,budroni'13,rand'16,bm'14,brierley'19,usha'13,alok'17,das'18,ku'18,titas'18,alok2,urbasi1} as well as application perspectives \cite{knee'12,robens'15,knee'16,ku'19,shayan'19,spee'20,Maity21}. (For a detailed survey,  see  \cite{emary'14,yearsley}). 

In the present work, we consider the framework of the prepare and measure
scenario (see, for instance, \cite{Armin'21}). We propose new inequalities for temporal correlations arising from sequential measurements and show that corresponding to every scenario of $n \mapsto 1$ RAC with SR there exists such an temporal inequality. Any quantum advantage of these communication tasks are implied by an associated violation of the derived temporal inequalities. These inequalities are asymmetric with respect to number of measurements. (For space-like separated correlations, asymmetric Bell inequalities were introduced to disprove the Peres conjecture \cite{Vertesi14}.) Moreover, as an immediate consequence of our result, maximal violation of these temporal inequalities provide the maximal success probability of the corresponding RAC with SR. In general,  maximal success probability of such RAC is  derived numerically for $n\geq 1$ with  unproven optimality  \cite{Ambainis_SR}. Moving on, we find an important application of our scheme in a cryptographically primitive task, {\it viz.} randomness generation. We show that any non-zero quantum advantage of RAC can be used for certifying randomness, while all previously proposed protocols for randomness generation based on RAC do not generate genuine randomness for any arbitrary success probability. \cite{rangen,qkd,randomness1}.

The plan of the paper is as follows. In section \ref{s2}, a preliminary discussion on $n \mapsto 1$ RAC is provided. In section \ref{s3}, with the aim of designing operational criteria for testing RAC,  temporal inequalities have been proposed for each $n \mapsto 1$ RAC. In section \ref{s4}, it has been shown that any non-zero quantum advantage of $n \mapsto 1$ RAC can be used to generate genuine randomness. Finally, section \ref{s5} is reserved for discussion on the results obtained in this paper along with some future directions.

%%%%%%%%%%%%%%%%%%%%%%%%%%%%%%%%%%%%%%%%%%%%%%%%%%%%%%%%%%%%%%%%%%%%%%%
%%%%%%%%%%%%%%%%%  up to here %%%%%%%%%%%%%%%%%%%%%%%%%%%%%%%%%
%%%%%%%%%%%%%%%%%%%%%%%%%%%%%%%%%%%%%%%%%%%%%%%%%%%%%%%%%%%%%%%%%

\section{Preliminaries} \label{s2}
%{\color{red}Leggett and Garg formalized two important macroscopic notions: \textbf{(A1)} {\it Macro-realism}(MR) and \textbf{(A2)} {\it Noninvasive measurability} (NIM). \textbf{MR} indicates that, in a system, if two states are macroscopically discernible, then there exists a macroscopically observable difference between them. These macroscopic observables would be considered to always have determinate values, at all times. On the other hand, \textbf{NIM} implies that, it is feasible, at least in principle,  to measure the state of a system, without having any impact on how it will evolve in the future.\\
%We derive a class of temporal inequalities based on these assumptions.}
Let us now discuss the idea of $n \mapsto 1$ RAC in some details. Consider a prepare and measure scenario where Alice depending on the $n$-bit input string given to her uniformly at random, implements a preparation procedure by encoding  the string in a qubit state $\rho_{x_1,...,x_n}$. The qubit is then sent to the measurement device held by Bob, who, upon receiving an input $y \in\{1,...,n\}$, implements a binary outcome measurement $B_{y}$ and reports the outcome $\beta \in\lbrace 0,1\rbrace$ as his output. Bob wins the game if he can perfectly guess the encoded bit sent by Alice. The average probability of winning is given by
\begin{align}\label{PnRAC}
\mathbb{F}_{{n \mapsto 1}} &= P(\beta_y= \text{y-th bit of Alice}) \nonumber \\
&= \frac{1}{n2^n}\sum_{x_1,...,x_n,y}P(\beta_y=x_y\mid x_1,...,x_n,y),
\end{align}
where the coefficient $\frac{1}{n2^n}$ appears due to normalization. Below we discuss some particular examples of $n \mapsto 1$ RAC and their explicit forms.

\textbf{$2 \mapsto 1$ RAC:} In a $2 \mapsto 1$ RAC, Alice has an input string consisting of two bits $x_1x_2\in\{00,01,10,11\}$ which she wants to send to Bob by encoding the bits in a qubit $\rho_{x_1x_2}$. Bob's task is to guess one of the bits (which is again chosen randomly) reliably. Therefore, after receiving an input $y \in\{1,2\}$, he performs a binary outcome measurement $B_{y}$ and reports the outcome $\beta \in\{0,1\}$ as his output. In this scenario, following Eq. \eqref{PnRAC} the average probability of perfectly guessing the bit is given by
\begin{align}\label{P2RAC}
\mathbb{F}_{{2 \mapsto 1}} = \frac{1}{8}\sum_{x_1,x_2,y}P(\beta_y=x_y\mid x_1,x_2,y).
\end{align}

Now consider the most general preparations and measurements as,
\begin{align}\label{pre2to1}
    \rho_{xx}&=\frac{1}{2}\left[ \mathbb{I}+ (-1)^x \hat{a_1}.\vec{\sigma}\right],~~\rho_{x\Bar{x}}=\frac{1}{2}\left[ \mathbb{I} +(-1)^x \hat{a_2}.\vec{\sigma}\right],
\end{align}
\begin{align}
    B_1&=\frac{1}{2}\left[ \mathbb{I}+ \left( -1\right) ^{b_1}\hat{b_1}.\vec{\sigma}\right],~~B_2=\frac{1}{2}\left[ \mathbb{I}+ \left( -1\right) ^{b_2}\hat{b_2}.\vec{\sigma}\right].
 \end{align}
Here, $\hat{a_i}$, $\hat{b_j}$ are the Bloch vectors denoting Alice’s and Bob’s measurement directions respectively and $\vec{\sigma}$ are the Pauli matrices.

Clearly, $\rho_{00}+\rho_{11} =\mathbb{I}$, $\rho_{01}+ \rho_{10} =\mathbb{I}$, $B_j^0 + B_j^1 =\mathbb{I}$ for $j=\{1,2\}$ and the notation $B_j^{b_j}$ denotes the eigenstate corresponding to the outcome $b_j$ of measurement $B_j$. The average success probability now can be written as
\begin{align}\label{rac2to1}
    \mathbb{F}_{2 \mapsto 1} &= \frac{1}{8}[\text{Tr}[\rho_{00}B^0_1+\rho_{00}B^0_2+\rho_{11}B^1_1+\rho_{11}B^1_2 \nonumber \\ 
    &~~~~+\rho_{01}B^0_1+\rho_{01}B^1_2+\rho_{10}B^1_1+\rho_{10}B^0_2]].
\end{align}
The maximum achievable value for the above quantity is  $\frac{1}{2}(1+\frac{1}{2})$ with classical strategy, whereas it can reach up to $\frac{1}{2}(1+\frac{1}{\sqrt2})$ if quantum strategy is used \cite{Ambainis_SR}.

\textbf{$3 \mapsto 1$ RAC:} In a $3 \mapsto 1$ RAC, a three-bit input string $x_1x_2x_3$ from the set $\{000,001,010,011,100,101,110,111\}$ is given to Alice uniformly at random. Alice then encodes the input string in a qubit $\rho_{x_1x_2x_3}$ and sends it to Bob. Bob upon receiving an input $y \in\{1,2,3\}$, implements a binary outcome measurement $B_{y}$ and reports the outcome $\beta \in\{0,1\}$ as his output. Following Eq. \eqref{PnRAC}, the average probability of winning can be calculated as
\begin{align}\label{P3RAC}
\mathbb{F}_{3 \mapsto 1} = \frac{1}{24}\sum_{x_1,x_2,x_3,y}P(\beta_y=x_y\mid x_1,x_2,x_3,y).
\end{align}

Let us now consider most general preparations 
\begin{align}\label{pre3to1}
    \rho_{xxx} &= \frac{1}{2}\left[ \mathbb{I}+ (-1)^x \hat{a_1}.\vec{\sigma}\right],\nonumber \\
    \rho_{xx\Bar{x}} &= \frac{1}{2}\left[ \mathbb{I}+ (-1)^x \hat{a_2}.\vec{\sigma}\right],\nonumber \\
    \rho_{x\Bar{x}x} &= \frac{1}{2}\left[ \mathbb{I}+ (-1)^x \hat{a_3}.\vec{\sigma}\right], \nonumber \\
    \rho_{x\Bar{x}\Bar{x}} &= \frac{1}{2}\left[ \mathbb{I}+ (-1)^x \hat{a_4}.\vec{\sigma}\right],
\end{align}
and measurements as,
\begin{align}
      B_1&=\frac{1}{2}\left[ \mathbb{I}+ \left( -1\right) ^{b_1}\hat{b_1}.\vec{\sigma}\right],\nonumber \\
      B_2&=\frac{1}{2}\left[ \mathbb{I}+ \left( -1\right) ^{b_2}\hat{b_2}.\vec{\sigma}\right],\nonumber \\
      B_3&=\frac{1}{2}\left[ \mathbb{I}+ \left( -1\right) ^{b_3}\hat{b_3}.\vec{\sigma}\right].
\end{align}

Clearly, $\rho_{000}+\rho_{111} =\mathbb{I}$, $\rho_{001}+\rho_{110} =\mathbb{I}$, $\rho_{010}+\rho_{101} =\mathbb{I}$, and $\rho_{011}+\rho_{100} =\mathbb{I}$. The average success probability can be written as
\begin{align}\label{rac3to1}
    \mathbb{F}_{3 \mapsto 1} &= \frac{1}{24}[\text{Tr}[\rho_{000}(B_1^0 +B_2^0 + B_3^0) + \rho_{001}(B_1^0+B_2^0+B_3^1) \nonumber \\
    &~~~~+ \rho_{010}(B_1^0+B_2^1+B_3^0)+\rho_{011}(B_1^0+B_2^1+B_3^1) \nonumber \\
    &~~~~+\rho_{100}(B_1^1+B_2^0+B_3^0)+\rho_{101}(B_1^1+B_2^0+B_3^1) \nonumber \\
    &~~~~+\rho_{110}(B_1^1+B_2^1+B_3^0)+ \rho_{111}(B_1^1+B_2^1+B_3^1)]].
\end{align}

Here, the average success probability, $\mathbb{F}_{3 \mapsto 1}$ can be achieved up to $\frac{1}{2}(1+\frac{1}{3})$ and $\frac{1}{2}(1+\frac{1}{\sqrt3})$, by classical and quantum strategies, respectively \cite{Ambainis_SR}.

\textbf{$4\mapsto 1$ RAC:} In a $4 \mapsto 1$ RAC, Alice has a four-bits
input string $x_1x_2x_3x_4$ which is given to her uniformly at random from the set $\{0000,0001,0010,0011,0100,0101,0110,0111,1000,1001,\\1010,1011,1100,1101,1110,1111\}$. She then implements a preparation procedure by encoding in a qubit $\rho_{x_1x_2x_3x_4}$ and sends to Bob. Bob upon receiving an input $y \in\{1,2,3,4\}$, implements a binary outcome measurement $B_{y}$ and reports the outcome $\beta \in\{0,1\}$ as his output. The average probability of winning is given by Eq. \eqref{PnRAC} as
\begin{align}\label{P4RAC}
\mathbb{F}_{4 \mapsto 1} = \frac{1}{64}\sum_{x_1,x_2,x_3,x_4,y}P(\beta_y=x_y\mid x_1,x_2,x_3,x_4,y).
\end{align}

Let us now consider most general preparations,
\begin{align}\label{pre4to1}
    \rho_{xxxx}&= \frac{1}{2}\left[ \mathbb{I}+ (-1)^x \hat{a_1}.\vec{\sigma}\right],\nonumber \\
    \rho_{xxx\Bar{x}}&= \frac{1}{2}\left[ \mathbb{I}+ (-1)^x \hat{a_2}.\vec{\sigma}\right],\nonumber \\
    \rho_{xx\Bar{x}x}&= \frac{1}{2}\left[ \mathbb{I}+ (-1)^x \hat{a_3}.\vec{\sigma}\right],\nonumber \\
    \rho_{xx\Bar{x}\Bar{x}}&= \frac{1}{2}\left[ \mathbb{I}+ (-1)^x \hat{a_4}.\vec{\sigma}\right],\nonumber \\
    \rho_{x\Bar{x}xx}&=  \frac{1}{2}\left[ \mathbb{I}+ (-1)^x \hat{a_5}.\vec{\sigma}\right],\nonumber \\
    \rho_{x\Bar{x}x\Bar{x}}&= \frac{1}{2}\left[ \mathbb{I}+ (-1)^x \hat{a_6}.\vec{\sigma}\right],\nonumber \\
    \rho_{x\Bar{x}\Bar{x}x}&= \frac{1}{2}\left[ \mathbb{I}+ (-1)^x \hat{a_7}.\vec{\sigma}\right],\nonumber \\
    \rho_{x\Bar{x}\Bar{x}\Bar{x}}&= \frac{1}{2}\left[ \mathbb{I}+ (-1)^x \hat{a_8}.\vec{\sigma}\right],
\end{align}
and measurements
\begin{align}
     B_1=\frac{1}{2}\left[ \mathbb{I}+ \left( -1\right) ^{b_1}\hat{b_1}.\vec{\sigma}\right],\nonumber \\
     B_2=\frac{1}{2}\left[ \mathbb{I}+ \left( -1\right) ^{b_2}\hat{b_2}.\vec{\sigma}\right],\nonumber \\
     B_3=\frac{1}{2}\left[ \mathbb{I}+ \left( -1\right) ^{b_3}\hat{b_3}.\vec{\sigma}\right],\nonumber \\
     B_4=\frac{1}{2}\left[ \mathbb{I}+ \left( -1\right) ^{b_4}\hat{b_4}.\vec{\sigma}\right].
\end{align}

Clearly, $\rho_{0000}+\rho_{1111} =\mathbb{I}$, $\rho_{0001}+\rho_{1110}  =\mathbb{I}$, $\rho_{0010}+\rho_{1101}  =\mathbb{I}$, $\rho_{0011}+\rho_{1100}  =\mathbb{I}$, $\rho_{0100}+\rho_{1011}  =\mathbb{I}$, $\rho_{0101}+\rho_{1010}  =\mathbb{I}$, $\rho_{0110}+\rho_{1001}  =\mathbb{I}$, and $\rho_{0111}+\rho_{1000}  =\mathbb{I}$.
Therefore the explicit form of the average success probability of $4 \mapsto 1$ RAC is calculated to be
\begin{align}\label{rac4to1}
    \mathbb{F}_{4 \mapsto 1} &= \frac{1}{64}[\text{Tr}[\rho_{0000}(B^0_1+B^0_2+B^0_3+B^0_4) +\rho_{0001}(B^0_1+B^0_2 \nonumber \\
    &+B^0_3+B^1_4)+\rho_{0010}(B^0_1+B^0_2+B^1_3+B^0_4)+\rho_{0011}(B^0_1 \nonumber \\
    &+B^0_2+B^1_3+B^1_4)+\rho_{0100}(B^0_1+B^1_2+B^0_3+B^0_4)+\rho_{0101} \nonumber \\
    &(B^0_1+B^1_2+B^0_3+B^1_4)+\rho_{0110}(B^0_1+B^1_2+B^1_3+B^0_4) \nonumber \\
    &+\rho_{0111}(B^0_1+B^1_2+B^1_3+B^1_4)+\rho_{1000}(B^1_1+B^0_2+B^0_3 \nonumber \\
    &+B^0_4)+\rho_{1001}(B^1_1+B^0_2+B^0_3+B^1_4)+\rho_{1010}(B^1_1+B^0_2 \nonumber \\ 
    &+B^1_3+B^0_4)+\rho_{1011}(B^1_1+B^0_2+B^1_3+B^1_4)+\rho_{1100}(B^1_1 \nonumber \\
    &+B^1_2+B^0_3+B^0_4)+\rho_{1101}(B^1_1+B^1_2+B^0_3+B^1_4)+\rho_{1110} \nonumber \\
    &(B^1_1+B^1_2+B^1_3+B^0_4)+\rho_{1111}(B^1_1+B^1_2+B^1_3+B^1_4)]].
\end{align}
For $4 \mapsto 1$ RAC, the exact value of classical and quantum average success probabilities are $\frac{1}{2}(1+\frac{1}{4})$ and $\frac{1}{2}(1+\frac{1+\sqrt3}{4\sqrt2})$ respectively \cite{Ambainis_SR}.

    \textbf{$n \mapsto 1$ RAC:} In a $n \mapsto 1$ RAC, Alice has an $n$-bits input string $x_1x_2x_3x_4\dots x_n$ which is given to her uniformly at random. She then implements a preparation procedure by encoding this string in a qubit denoted by $\rho_{x_1x_2x_3x_4\dots x_n}$ and sends it to Bob. Bob upon receiving an input $y \in\{1,2,3,\dots,n\}$, implements a binary outcome measurement $B_{y}$ and reports the outcome $\beta \in\{0,1\}$ as his output. The average probability of winning is given by Eq. \eqref{PnRAC}. 
For $n \mapsto 1$ RAC, there are $2^{n-1}$ preparations and $n$ measurements. 
Let us now consider the most general preparations,
\begin{align}\label{prento1}
    \rho_{xxx\dots xx}&= \frac{1}{2}\left[ \mathbb{I}+ (-1)^x \hat{a}_1.\vec{\sigma}\right],\nonumber \\
    \rho_{xxx\dots x\Bar{x}}&= \frac{1}{2}\left[ \mathbb{I}+ (-1)^x \hat{a}_2.\vec{\sigma}\right],\nonumber \\
    \rho_{xxx\dots\Bar{x}x}&= \frac{1}{2}\left[ \mathbb{I}+ (-1)^x \hat{a}_3.\vec{\sigma}\right],\nonumber \\
    \noalign{\centering$\vdots$}
    \rho_{xx\Bar{x}\dots \Bar{x}\Bar{x}}&= \frac{1}{2}\left[ \mathbb{I}+ (-1)^x \hat{a}_{2^{n-1}-1}.\vec{\sigma}\right],\nonumber \\
    \rho_{x\Bar{x}\Bar{x}\dots \Bar{x}\Bar{x}}&= \frac{1}{2}\left[ \mathbb{I}+ (-1)^x \hat{a}_{2^{n-1}}.\vec{\sigma}\right],
\end{align}
and measurements
\begin{align*}
     B_1=\frac{1}{2}\left[ \mathbb{I}+ \left( -1\right) ^{b_1}\hat{b}_1.\vec{\sigma}\right],\nonumber \\
     B_2=\frac{1}{2}\left[ \mathbb{I}+ \left( -1\right) ^{b_2}\hat{b}_2.\vec{\sigma}\right],\nonumber \\
     B_3=\frac{1}{2}\left[ \mathbb{I}+ \left( -1\right) ^{b_3}\hat{b}_3.\vec{\sigma}\right],\nonumber \\
     \noalign{\centering$\vdots$}
     B_n=\frac{1}{2}\left[ \mathbb{I}+ \left( -1\right) ^{b_n}\hat{b}_n.\vec{\sigma}\right].
\end{align*}

Clearly, $\rho_{m}+\rho_{\Bar{m}} =\mathbb{I}$,
where, $m$ are the elements of the n-bit string set.
The exact form of the average success probability of $n \mapsto 1$ RAC can be calculated as
\begin{align}\label{racnto1}
    \mathbb{F}_{n \mapsto 1} &= \frac{1}{n2^n}[\text{Tr}[\rho_{00\dots 00}(B^0_1+B^0_2+\dots +B^0_{n-1}+B^0_n) + \nonumber\\
    &\rho_{00\dots 01}(B^0_1+B^0_2+\dots +B^0_{n-1}+B^1_n) +\rho_{00\dots 10}(B^0_1 \nonumber\\
    &+B^0_2+\dots +B^1_{n-1}+B^0_n)+\dots +\rho_{11\dots 10} (B^1_1+B^1_2\nonumber\\
    &+\dots +B^1_{n-1}+B^0_n)+\rho_{11\dots 11}(B^1_1+B^1_2+\dots +\nonumber\\
    &B^1_{n-1}+B^1_n)]].
\end{align}
The average success probability for $n \mapsto 1$ RAC (with SR) using best classical strategy is known to be $\frac{1}{2}(1+\frac{1}{n})$ \cite{Ambainis_SR}.

\section{Temporal inequalities associated with the Random access codes} \label{s3}
We would now like to present a temporal inequality corresponding to each $n \mapsto 1$ RAC. To derive such temporal inequalities, we use the assumptions of realism and noninvasive measurability. The term `realism' implies {\it ``at any instant, irrespective of any measurement, a system is definitely in any one of the available states such that all its observable properties have definite values"}. On the other hand, the term `noninvasive measurability' assures that {\it ``it is possible, in principle, to determine which of the states the system is in, without affecting the state itself or the system’s subsequent evolution"}. It can be shown that under these two assumptions the joint probability  distribution get factorized at the ontological level \cite{LG'85,emary'14,yearsley,kofler'13}. Mathematically, 
\begin{equation}
   P(a_i,b_j \mid A_i,B_j) = \int_\lambda \rho (\lambda) P(a_i \mid A_i, \lambda) P(b_j\mid B_j, \lambda ) 
\end{equation}
with $\lambda$ being the hidden variable. Based on this macro-realistic definition of classicality, later we derive temporal inequalities corresponding to each $n \mapsto 1$ RAC and establish that violation of such inequalities  implies non-classical temporal correlation. The underlying experimental setup for both the scenarios are the same, as will be clear from the following description.  

 In a temporal scenario, temporal correlations are obtained by measuring a single system sequentially at different instants of time. In each run of the experiment, two sequential measurements are performed on an identically prepared initial state. We assume that the first measurement is performed by Alice whereas the second one is performed by Bob. If Alice's measurement is thought of as playing the role of preparation, then it is not very difficult to realize the similarity between RACs and macrorealistic inequalities. This observation is 
further substantiated by formulating the relevant inequalities which build the connection quantitatively.   Suppose the measurements performed by Alice and the corresponding outcomes are denoted by $A_i$ and $a_i$ respectively. Similarly, the measurements performed by Bob are denoted by $B_j$ with corresponding outcome $b_j$. All the measurements performed by both the parties are considered to be dichotomic {\it i.e.}, $a_i,b_j \in \lbrace 0,1 \rbrace$. 

Initially, the state on which Alice performs her measurement is denoted by $\rho_{in}$. The correlation between Alice's and Bob's measurement outcome depends on $\rho_{in}$. In the present analysis we took a maximally mixed state, i.e., $\rho_{in}= \mathbb{I}/2$, for reasons that will be clear
later. It may be noted here that other states may be chosen for which the same maximum violation for our temporal inequalities can be achieved using different Bloch vectors ($\hat{a_i}$, $\hat{b_j}$). However, even if the directions of Alice’s and Bob’s measurements are changed, the maximum violation will remain the same.

Let the probability of obtaining outcome $a_i$ and $b_j$ be denoted by $P(a_i,b_j \mid A_i,B_j)$, when Alice measures $A_i$ at time $t_i$ and Bob measures $B_j$ at some later instant $t_j$ respectively and $a_i, b_j\in\{0,1\}$. Let us denote, $A_i^{a_i}$, $B_j^{b_j}$ as projectors so that $\sum_{a_i}A_i^{a_i}=\mathbb{I}, \sum_{b_j}B_j^{b_j}=\mathbb{I}$. Now, following the standard procedure, the joint probability distribution can be obtained using Bayes' rule as,
\begin{align}\label{pij}
&P(a_i,b_j \mid A_i,B_j) = P(a_i \mid A_i) P(b_j\mid a_i, A_i,B_j ) \nonumber \\
&= \text{Tr}\left[ A_i^{a_i} \rho_{in}\right] \text{Tr}\left[ B_j^{b_j} \frac{A_i^{a_i}\rho_{in}A_i^{a_i \dagger}}{\text{Tr}\left[ A_i^{a_i}\rho_{in}A_i^{a_i \dagger}\right]}\right] .
\end{align}
With this joint probability distribution, the two-time correlation is defined as,
\begin{equation}\label{cij}
C_{ij} =\sum_{a_i,b_j} (-1)^{a_i\oplus b_j} P(a_i,b_j \mid A_i,B_j),
\end{equation}
where $\oplus$ denotes addition modulo 2. In the subsections below we present the temporal inequalities corresponding to cases of $2 \mapsto 1$, $3 \mapsto 1$
and $4 \mapsto 1$ RACs.  These inequalities are derived with a close look at the success probabilities of the corresponding RAC games. The general form of the temporal inequality for
$n \mapsto 1$ RAC, i.e., $\mathcal{K}_{n \mapsto 1}$ is provided in the Appendix-\eqref{appendixb}.

As mentioned earlier, in our temporal scenario, the initially prepared state is considered to be $\mathbb{I}/2$. Since we are interested in finding the maximum violation of the temporal inequality $\mathcal{K}_{n \mapsto 1}$, without loss of generality we can stick to projective measurements only. Consider the general form of the measurements on Alice's side to be
\begin{equation}
    A_i^{a_i} =\frac{1}{2}\left[ \mathbb{I}+ (-1)^{a_i} \hat{a_i}.\vec{\sigma}\right],
\end{equation}
where $a_i$ represents the outcome corresponding the measurement $A_i$ and $\hat{a}_i$ represents the direction along which the $A_i$ measurement is being performed.
On the other hand the general form of the measurements performed on Bob's side can be considered to be
\begin{equation}
    B_j^{b_j} =\frac{1}{2}\left[ \mathbb{I}+ (-1)^{b_j} \hat{b_j}.\vec{\sigma}\right],
\end{equation}
where $b_j$ represents the outcome corresponding to measurement $B_j$ and $\hat{b}_j$ represents the direction along which $B_j$ measurement is being performed. In the subsections below we state the explicit form of the quantum strategy for which maximum quantum violation of the temporal inequality $\mathcal{K}_{n \mapsto 1}$ RAC is achieved.

\subsection{Temporal inequality for $2 \mapsto 1$ RAC} Now, to derive a temporal inequality corresponding to the $2 \mapsto 1$ RAC, let us first assume that Alice and Bob have two choices of binary measurements, say, $\{A_1,A_2\}$ and $\{B_1,B_2\}$ to perform in each run and $\rho_{in}= \frac{\mathbb{I}}{2}$. Let us now consider the following quantity in terms of the above correlators as,
\begin{equation}\label{k2to1}
\mathcal{K}_{2 \mapsto 1} = C_{11} + C_{21} + C_{12} - C_{22}.
\end{equation}

Following Eq. \eqref{cij}, we calculate the $C_{11}$ term explicitly:
\begin{align}
C_{11} =& P(0,0|A_1,B_1)+ P(1,1|A_1,B_1)- \nonumber \\
&P(0,1|A_1,B_1) -P(1,0|A_1,B_1) \nonumber \\
=&\frac{1}{2}\text{Tr}[A_1^0B_1^0 + A_1^1B_1^1 - A_1^0B_1^1- A_1^1B_1^0] \nonumber\\
=&\frac{1}{2}\text{Tr}[A_1^0B_1^0 +A_1^1B_1^1 - A_1^0(\mathbb{I}-B_1^0)- A_1^1(\mathbb{I}-B_1^1)] \nonumber \\
=&\text{Tr}[A^0_1B_1^0+A_1^1B_1^1]-1. \nonumber
\end{align}
where, $A_i^{a_i}$ represents the eigenstate corresponding to the outcome $a_i \in \lbrace 0,1\rbrace$ of the measurement $A_i$ and similarly for Bob. Here, the second equality can be derived using Eq.\eqref{pij}, {\it i.e.,} by evaluating all the probability terms explicitly, and the third equality follows from the fact that two eigenstate corresponding to the same dichotomic measurement add up to unity, {\it i.e.,} $A_i^0+A_i^1=\mathbb{I}$ and $B_j^0+B_j^1=\mathbb{I}$ for all $i,j$.

Similarly, the other terms can be evaluated as
\begin{align}
C_{12} &=\text{Tr}[ A_1^0B^0_2+A_1^1B^1_2]-1, \nonumber \\
C_{21} &=\text{Tr}[A_2^0B^0_1+A_2^1B^1_1]-1, \nonumber \\
C_{22} &=-\text{Tr}[A_2^1B^0_2+A_2^0B^1_2]+1. \nonumber
\end{align}
Hence, the expression for $\mathcal{K}_{2 \mapsto 1}$ becomes,
\begin{align}
    \mathcal{K}_{2 \mapsto 1} &= C_{11} + C_{21} + C_{12} - C_{22} \nonumber \\
   &= \text{Tr}[A^0_1B_1^0+A_1^1B_1^1+A_1^0B^0_2+A_1^1B^1_2+A_2^0B^0_1 \nonumber \\
   &~~~~~+A_2^1B^1_1+A_2^1B^0_2+A_2^0B^1_2]-4.
\end{align}
It may be noted here that the eigenstate of the measurements performed by Alice $A_i^{a_i}$ is the same as that of the preparations considered in Eq.\eqref{pre2to1}. Now, comparing the above equation with  Eq.\eqref{rac2to1}, one  obtains
\begin{equation}
    \mathcal{K}_{2 \mapsto 1} =8(\mathbb{F}_{2 \mapsto 1}-\frac{1}{2}).
\end{equation}

Conversely, $\mathbb{F}_{2 \mapsto 1}=\frac{1}{2}+\frac{1}{8}\mathcal{K}_{2 \mapsto 1}$.

It can be shown that the maximum quantum violation of the temporal inequality $\mathcal{K}_{2 \mapsto 1}$ corresponding to $2 \mapsto 1$ RAC can be achieved up to $2.828$ for the following sets of measurements:
\begin{align}
    \hat{a}_1 &= \frac{1}{\sqrt{2}}(1,1,0), \nonumber \\
    \hat{a}_2 &= \frac{1}{\sqrt{2}}(1,-1,0),
\end{align}
and
\begin{align}
    \hat{b}_1 &= (1,0,0), \nonumber \\
    \hat{b}_2 &= (0,1,0).
\end{align}
One can see that the strategy to reach the maximum violation of the temporal inequality $\mathcal{K}_{2 \mapsto 1}$ with initially prepared state $\mathbb{I}/2$ is the same with that of the quantum strategy for which the maximum success probability for $2 \mapsto 1$ RAC is achieved.

Note further, a noninvasive-realist bound for the term $\mathcal{K}_{2 \mapsto 1}$ was derived to be $2$ (See, Appendix-\eqref{ap1}). One can see from this  relation that whenever the value of the term $\mathcal{K}_{2 \mapsto 1}$ falls below $2$, the success probability of the $2 \mapsto 1$ random access code also falls below $\frac{3}{4}$ which is the maximum probability of success with classical strategy. 
Moreover, the maximum success probability of $2 \mapsto 1$ RAC reaches $\frac{1}{2}(1+\frac{1}{\sqrt{2}})$ whenever the maximal qubit strategy of $\mathcal{K}_{2 \mapsto 1}$ reaches $2\sqrt{2}$, and vice versa.  Therefore, any quantum advantage of $2 \mapsto 1$ RAC implies a violation of the corresponding macro-realist model.

\subsection{Temporal inequality for $3 \mapsto 1$ RAC} To derive a temporal inequality analogous to $3 \mapsto 1$ RAC, we need to consider four measurements on Alice's side say $\{A_1,A_2,A_3,A_4\}$ and three measurements on Bob's side say $\{B_1,B_2,B_3\}$. In each run of the experiment, Alice  performs one out of the four dichotomic measurements on an initially prepared input state, $\rho_{\text{in}} = \frac{\mathbb{I}}{2}$, and then Bob  implements one out of the three possible measurements on the post-measurement state of Alice. In this way let us define the following quantity in terms of the correlators \eqref{cij} as,
\begin{align}\label{k3to1}
\mathcal{K}_{3 \mapsto 1} &= C_{11} + C_{12} + C_{13} + C_{22} + C_{21} - C_{23} + C_{31} - C_{32} \nonumber \\ 
&~~~~~+ C_{33} + C_{41} - C_{42} - C_{43}.
\end{align}
Following Eq. \eqref{cij} and after some straightforward calculations, one can obtain the general form of the correlators as
\begin{align}\label{cijg}
    C_{ij}&=(-1)^{a_i} [\text{Tr}[A_i^{a_i}B_j^0+A_i^{\Bar{a_i}}B_j^1]-1].
\end{align}
Here, $i$ and $j$ denotes Alice's and Bob's measurement indices, respectively. The outcome of Alice's measurement $A_i^{a_i}$ are denoted as $a_i$ and $\Bar{a_i}$ represents the complement of $a_i$.

Therefore, the term $\mathcal{K}_{3 \mapsto 1}$ becomes,
\begin{align}\label{k31}
\mathcal{K}_{3 \mapsto 1} =\sum_{j=1}^3\sum_{i=1}^4 (-1)^{a_i} [\text{Tr}[A_i^{a_i}B_j^0+A_i^{\Bar{a_i}}B_j^1]-1]\nonumber \\
= \sum_{j=1}^3\sum_{i=1}^4 (-1)^{a_i} \text{Tr}[A_i^{a_i}B_j^0+A_i^{\Bar{a_i}}B_j^1]-12.
\end{align}

Taking the eigenstate of Alice's measurement $A_i^{a_i},~i\in \{1,2,3,4\}$ as that of the preparations for $3 \mapsto 1$ RAC, {\it i.e.}, Eq.\eqref{pre3to1} we obtain the success probability of $3 \mapsto 1$ RAC to be
\begin{align}
   \mathcal{K}_{3 \mapsto 1} = 24\mathbb{F}_{3 \mapsto 1}-12
\end{align}
or equivalently, $\mathbb{F}_{3 \mapsto 1} = \frac{1}{2}+\frac{1}{24}\mathcal{K}_{3 \mapsto 1}$.

It can be shown that the maximum violation of the temporal inequality $\mathcal{K}_{3 \mapsto 1}$ corresponding to the $3 \mapsto 1$ RAC can be achieved up to $6.928$ for the following measurement settings:
\begin{align}
    \hat{a}_1 &= \frac{1}{\sqrt{3}}(1,1,1), \nonumber \\
    \hat{a}_2 &= \frac{1}{\sqrt{3}}(1,1,-1), \nonumber \\
    \hat{a}_3 &= \frac{1}{\sqrt{3}}(1,-1,1), \nonumber \\
    \hat{a}_4 &= \frac{1}{\sqrt{3}}(1,-1,-1).
\end{align}
and
\begin{align}
    \hat{b}_1 &= (1,0,0), \nonumber \\
    \hat{b}_2 &= (0,1,0), \nonumber \\
    \hat{b}_3 &= (0,0,1).
\end{align}
This is again the same strategy for which maximum quantum success probability of $3 \mapsto 1$ RAC is achieved.

A noninvasive-realist bound for the above quantity  $\mathcal{K}_{3 \mapsto 1}$ is derived in Appendix-\eqref{ap2} to be $4$. One can see that when $\mathcal{K}_{3 \mapsto 1}=4$, the success probability $\mathbb{F}_{3 \mapsto 1}$ reaches $\frac{2}{3}$ which is the best classical strategy to win the $3 \mapsto 1$ RAC. On the other hand, with the maximal qubit strategy, $\mathcal{K}_{3 \mapsto 1}$ can however achieve value up to $6.928$, and for this the success probability $\mathbb{F}_{3 \mapsto 1}$ to win $3 \mapsto 1$ RAC reaches up to $\frac{1}{2}(1+\frac{1}{\sqrt{3}})$. This is again the maximum success probability of winning $3 \mapsto 1$ RAC using quantum strategy. Therefore, any violation of $3 \mapsto 1$ RAC again does not possess any macro-realist description.

\subsection{Temporal inequality for $4 \mapsto 1$ RAC} To derive a temporal inequality corresponding to $4 \mapsto 1$ RAC, Alice and Bob need to perform eight and four measurements respectively in their respective parts. In each run of the experiment Alice performs one out of the eight dichotomic measurements on an initially prepared input state, $\rho_{\text{in}} = \frac{\mathbb{I}}{2}$, and Bob performs one out of the four dichotomic measurements on the post measurement state of Alice. Let us now consider the following quantity, consisting of thirty-two correlators given by,
\begin{align}\label{k4to1}
\mathcal{K}_{4 \mapsto 1} &= C_{11} + C_{12} + C_{13} + C_{14} + C_{21} + C_{22} + C_{23} - C_{24} \nonumber \\
&+ C_{31} + C_{32} - C_{33} + C_{34} + C_{41} + C_{42} - C_{43} - C_{44} \nonumber \\
&+ C_{51} - C_{52} + C_{53} + C_{54} + C_{61} - C_{62} + C_{63} - C_{64} \nonumber \\
&+ C_{71} - C_{72} - C_{73} + C_{74} + C_{81} - C_{82} - C_{83} - C_{84}.
\end{align}
Now, the explicit form of the correlators $C_{ij}$ can be evaluated directly from  Eq.\eqref{cijg}. Therefore, the term $\mathcal{K}_{4 \mapsto 1}$ reduces to
\begin{align}\label{k41}
\mathcal{K}_{4 \mapsto 1} =\sum_{j=1}^4\sum_{i=1}^8 (-1)^{a_i} [\text{Tr}[A_i^{a_i}B_j^0+A_i^{\Bar{a_i}}B_j^1]-1]\nonumber \\
= \sum_{j=1}^4\sum_{i=1}^8 (-1)^{a_i} \text{Tr}[A_i^{a_i}B_j^0+A_i^{\Bar{a_i}}B_j^1]-32
\end{align}
with the symbols having usual meanings. Now comparing the eigenstate of Alice's measurement $A_i^{a_i},~i\in \{1,2,...,8\}$ with that of the preparations for the $4 \mapsto 1$ {\it i.e.}, Eq. \eqref{pre4to1}, we obtain
\begin{equation}
    \mathcal{K}_{4 \mapsto 1} = 64(\mathbb{F}_{4 \mapsto 1}-\frac{1}{2}).
\end{equation}
Equivalently, $\mathbb{F}_{4 \mapsto 1} = \frac{1}{2}+\frac{1}{64}\mathcal{K}_{4 \mapsto 1}$.

The maximum quantum violation of the temporal inequality $\mathcal{K}_{4 \mapsto 1}$ can be achieved up to $15.454$ for the following measurement settings:
\begin{align}
    \hat{a}_1 &= \frac{1}{\sqrt{6}}(1,1,2), \nonumber \\
    \hat{a}_2 &= \frac{1}{\sqrt{6}}(1,1,-2), \nonumber \\
    \hat{a}_3 &= \frac{1}{\sqrt{6}}(1,-1,2), \nonumber \\
    \hat{a}_4 &= \frac{1}{\sqrt{6}}(1,-1,-2), \nonumber \\
    \hat{a}_5 &= \frac{1}{\sqrt{6}}(\sqrt{3},\sqrt{3},0), \nonumber \\
    \hat{a}_6 &= \frac{1}{\sqrt{6}}(\sqrt{3},-\sqrt{3},0), \nonumber \\
    \hat{a}_7 &= \frac{1}{\sqrt{6}}(\sqrt{3},\sqrt{3},0), \nonumber \\
    \hat{a}_8 &= \frac{1}{\sqrt{6}}(\sqrt{3},-\sqrt{3},0).
\end{align}
and
\begin{align}
    \hat{b}_1 &= (1,0,0), \nonumber \\
    \hat{b}_2 &= (0,1,0), \nonumber \\
    \hat{b}_3 &= (0,0,1), \nonumber \\
    \hat{b}_4 &= (0,0,1). 
\end{align}
This is again the same strategy for which maximum success probability of $4 \mapsto 1$ RAC is achieved.

A noninvasive-realist bound for $\mathcal{K}_{4 \mapsto 1}$ is derived in Appendix-\eqref{ap3} to be $8$. In addition, the best classical strategy to win the $4 \mapsto 1$ RAC is $\frac{5}{8}$. One can see from the above relation that whenever the $\mathcal{K}_{4 \mapsto 1}$ rises above $8$ there is no macro-realist model, and only in this case the quantum advantage of $4 \mapsto 1$ RAC
can  be obtained. The term $\mathcal{K}_{4 \mapsto 1}$ can reach up to $15.454$ with maximal qubit strategy  which also matches with the maximum average success probability, $\mathbb{F}_{4 \mapsto 1} = 0.741$.

\subsection{Temporal inequality for $n \mapsto 1$ RAC}
Let us now derive a temporal inequality corresponding to $n \mapsto 1$ RAC. To do so we need to consider $2^{n-1}$ measurements on Alice's side say $\{A_1,A_2,A_3,\dots,A_{2^{n-1}}\}$ and n measurements on Bob's side say $\{B_1,B_2,\dots,B_n\}$. At first, Alice  performs one out of the $2^{n-1}$ dichotomic measurements on an initially prepared input state, $\rho_{\text{in}} = \frac{\mathbb{I}}{2}$ and then, Bob  performs one out of the $n$ possible measurements on the post measurement state of Alice. Finally, they evaluate the quantity $\mathcal{K}_{n \mapsto 1}$, represented in terms of the correlators \eqref{cij} as,
\begin{align}\label{kn}
\mathcal{K}_{n \mapsto 1} &= C_{11} + C_{12} + C_{13} +\dots + C_{1n} + C_{21} + C_{22} +\nonumber\\
&C_{23} +\dots - C_{2n} + C_{31} + C_{32} + C_{33} +\dots +C_{3n} \nonumber\\
&+\dots +C_{{(2^{n-1}-1)}1} - C_{{(2^{n-1}-1)}2} - C_{{(2^{n-1}-1)}3} \nonumber\\
&+\dots +C_{{(2^{n-1}-1)}n} +\dots +C_{2^{n-1}1} - C_{2^{n-1}2} \nonumber\\
&- C_{2^{n-1}3} -\dots
- C_{2^{n-1}n}.
\end{align}
It might be noted here that some of the correlators will contain negative sign. This is because Alice's and Bob's measurements are anti-correlated for those particular terms.
Now, the explicit form of the correlators can be written from Eq.\eqref{cijg}.

Therefore, the term $\mathcal{K}_{n \mapsto 1}$ becomes,
\begin{align}\label{kn1}
\mathcal{K}_{n \mapsto 1} =\sum_{j=1}^n\sum_{i=1}^{2^{n-1}} (-1)^{a_i} [\text{Tr}[A_i^{a_i}B_j^0+A_i^{\Bar{a_i}}B_j^1]-1]\nonumber \\
= \sum_{j=1}^n\sum_{i=1}^{2^{n-1}} (-1)^{a_i} \text{Tr}[A_i^{a_i}B_j^0+A_i^{\Bar{a_i}}B_j^1]-n2^{n-1}.
\end{align}

Now if we take the eigenstate of Alice's measurement $A_i^{a_i},~i\in \{1,2,\dots,2^{n-1}\}$ as that of the preparations for $n \mapsto 1$ RAC, {\it i.e.}, Eq.\eqref{prento1}, we obtain the success probability of $n \mapsto 1$ RAC to be
\begin{align}\label{knto1}
   \mathcal{K}_{n \mapsto 1} = n2^n\mathbb{F}_{n \mapsto 1}-n2^{n-1}
\end{align}
or equivalently, $\mathbb{F}_{n \mapsto 1} = \frac{1}{2}+\frac{1}{n2^n}\mathcal{K}_{n \mapsto 1}$.

It can be checked that when $\mathcal{K}_{n \mapsto 1}=2^{n-1}$, the success probability $\mathbb{F}_{n \mapsto 1}$ reaches $\frac{1}{2}(1+\frac{1}{n})$ which is the best classical strategy to win the $n \mapsto 1$ RAC. In Appendix-\eqref{apn} we  derive the noninvasive-realist bound for the temporal inequality corresponding to the $n \mapsto 1$ RAC which turns out to be $2^{n-1}$. Now, from the relation between the average success probability $\mathbb{F}_{n \mapsto 1}$ and corresponding temporal inequality $\mathcal{K}_{n \mapsto 1}$, we can conclude that any non-zero quantum advantage of general $n \mapsto 1$ RAC necessitates non-classical temporal correlation. On the basis of the above investigation, following the main idea stated at the beginning of this section, we clearly summarize our first main result, given bellow.

$\bullet$ {\it \textbf{ Result 1:-} } For every $n \mapsto 1$ RAC with SR, there exists a temporal inequality where Alice, the first observer has $2^{n-1}$ measurement settings and Bob who measures later has $n$ measurement settings.
The maximum success probability of each $n \mapsto 1$ RAC with best classical strategy is related to the maximum noninvasive-realist bound, and
any non-zero quantum advantage of a $n \mapsto 1$ RAC translates to the violation of the corresponding temporal inequality. 

 Moreover, based on our analysis for the cases of $2 \mapsto 1$, $3 \mapsto 1$  
 and $4 \mapsto 1$ RACs, we can further make the following conjecture. If the maximum success probability of $n \mapsto 1$ RAC occurs with a set of encoding states for Alice and decoding measurements for Bob, then maximal violation of the corresponding temporal inequality is obtained with Alice measuring observables whose eigenstates are exactly the encoded states and Bob's measurements are decoding observables with the initially prepared input state $\mathbb{I}/2$.

%%%%%%%%%%%%%%%%%%%%%%%%%%%%%%%%%%%%%%%%%%%%%%%%%%%%%%%%%%%%%%%%%%%%%%%
%%%%%%%%%%%%%%%  up to here %%%%%%%%%%%%%%%%%%%%%%%%%%%%%%%
%%%%%%%%%%%%%%%%%%%%%%%%%%%%%%%%%%%%%%%%%%%%%%%%%%%%%%%%%%%%%%%%%%

\section{Certifying true randomness}  \label{s4}
For the purpose of certifying randomness we describe here an alternative derivation of temporal inequalities, instead of the one based on realism and noninvasive measurability.  This alternative derivation was proposed based on some operational assumptions which can be tested in a real experiment. In this alternative derivation the pertaining assumptions are {\it no signaling in time} (NSIT) and {\it predictability} \cite{kofler'13,rand'16,Maity'21}. The NSIT condition states that the measurement statistics are not influenced by the earlier measurements, or mathematically, $P(b_j|B_j)=P(b_j|A_i,B_j)  ~\forall A_i,B_j,b_j$ \cite{kofler'13}. On the other hand  a model is said to be predictable if $P(a_i,b_j|A_i,B_j) \in \{0,1 \} ~ \forall a_i,b_j,A_i,B_j$ \cite{eric}.

Now, if $\lambda$ denotes some classical variable at the ontological level, then in order to predict the experimental results at the operational level one needs to integrate over all $\lambda$, {\it i.e.}, $p(a_i,b_j|A_i,B_j)= \int_{\lambda} d\lambda p(\lambda)p(a_i,b_j|A_i,B_j,\lambda)$. Note that a crucial step to derive the temporal inequality is to show that the probability distribution at ontological level gets factorised, { i.e.\it},  
\begin{equation}
    p(a_i,b_j|A_i,B_j,\lambda)=p(a_i|A_i,\lambda)p(b_j|B_j,\lambda)
\end{equation}

Now, using predictability one can write $p(a_i,b_j|A_i,B_j,\lambda)=p(a_i,b_j|A_i,B_j)$ as further conditioning does not change the deterministic probability distribution. Using Bayes' rule one can write the probability distribution $p(a_i,b_j|A_i,B_j) = p(a_i|A_i,B_j,b_j)p(b_j|A_i,B_j)$. Now, from the NSIT conditions one  has $p(b_j|A_i,B_j)=p(b_j|B_j)$. Also, from a physically reasonable perspective, it is broadly accepted that a later measurement cannot influence the past measurement result, and hence $p(a_i|A_i,B_j,b_j)=p(a_i|A_i)$. Since, at the ontological level $p(a_i|A_i,\lambda) = p(a_i|A_i)$ and $p(b_j|B_j,\lambda)=p(b_j|B_j)$, the probability distribution can be written in a factorized form $p(a_i,b_j|A_i,B_j,\lambda)=p(a_i|A_i,\lambda)p(b_j|B_j,\lambda)$. Therefore, 
\begin{equation}
    \text{NSIT} \wedge \text{predictability} \implies \text{factorizability} 
\end{equation}
or 
\begin{equation}
   \neg ~\text{factorizability} \wedge \text{NSIT} \implies  \neg ~\text{predictability}.
\end{equation}
Hence, if we consider a set of probability distributions which satisfies the NSIT conditions but does not fulfill the conditions for factorizability, then it is sure that predictability must be violated. In other words, if for a set of probability distributions the NSIT conditions hold and the temporal inequality is violated simultaneously, then predictability must not hold. Let us quantify this randomness by min-entropy $H_{\infty}(X)$ which captures the associated randomness that a particular distribution $X$ contains. Therefore, any probability distribution $P(a_i,b_j|A_i,B_j)$ will not be predictable and hence some genuine randomness must be associated with that probability distribution  \cite{min-entropy}. For some distribution $P(a_i,b_j|A_i,B_j)$, the min-entropy is defined as 
\begin{align}
    H_{\infty}(a_i,b_j|A_i,B_j) &= -\log_2 [\text{max}_{a_i,b_j} P(a_i,b_j|A_i,B_j)] \nonumber \\
    &= \text{min}_{a_i,b_j} [ -\log_2 [ P(a_i,b_j|A_i,B_j)]].
\end{align}

To calculate the randomness associated with the $\mathcal{K}_{n \mapsto 1}$, we need to find the maximum probability distribution $P(a_i,b_j|A_i,B_j)$ corresponding to some violation of this inequality. In other words, we need to solve the following optimization problem \cite{randomness}. 
\begin{align}\label{abc}
    P^*(a_i,b_j|A_i,B_j) &= \text{max} P(a_i,b_j|A_i,B_j) \nonumber \\
     & \text{constraints to}~ \mathcal{K}_{n \mapsto 1} = \mathcal{K}^{MR}_{n \mapsto 1}+ \epsilon , \nonumber \\
     & P(a_i,b_j|A_i,B_j) \geq 0, \nonumber \\
     & \sum_{a_i,b_j} P(a_i,b_j|A_i,B_j) =1 ~~\forall A_i,B_j, \nonumber \\
     &\text{and} ~P(a_i,b_j|A_i,B_j) ~\text{satisfy NSIT}
\end{align}
where $P^*(a_i,b_j|A_i,B_j)$ denotes the maximized value of $P(a_i,b_j|A_i,B_j)$  and $\mathcal{K}^{MR}_{n \mapsto 1}$ is the MR bound of the temporal inequality $\mathcal{K}_{n \mapsto 1}$ corresponding to $n \mapsto 1$ RAC. We use linear programming to solve this optimization problem. The parameters $\alpha$ and $\beta$ are chosen in such a way that the inequality  maintain its linear form. By putting some boundary conditions, $\alpha$ and $\beta$ are calculated for each $\mathcal{K}_{n \mapsto 1}$ so that $P^*(a_i,b_j|A_i,B_j) \leq \alpha \mathcal{K}_{n \mapsto 1} + \beta$.
Here, it may be noted that for any particular $n\mapsto 1$ RAC, if $\alpha$ and $\beta$ depend on $a_i,b_j,A_i,B_j$, then assuming $P^*(a_i,b_j|A_i,B_j) \leq \alpha \mathcal{K}_{n \mapsto 1} + \beta$ is not consistent, since we chose $\alpha$ and $\beta$ to be some constant so that the  linear form of the inequality can be maintained. However, in case of a different $n\mapsto 1$ RAC, this $\alpha$ and $\beta$ in general depends on the inequality corresponding to $\mathcal{K}_{n \mapsto 1}$ as well as on $P^*(a_i,b_j|A_i,B_j)$. Therefore, in general it may depend on  $a_i,b_j,A_i,B_j$. A more detailed discussion on the choice of $\alpha$ and $\beta$ (for the Bell-scenario) is provided explicitly in Ref.\cite{randomness}.

Note for example, in the case of $2 \mapsto 1$ RAC, the MR or classical bound of $\mathcal{K}_{2 \mapsto 1}$ is $2$. Now, $\mathcal{K}_{2 \mapsto 1} = 2+ \epsilon$ (with $\epsilon > 0$)  implies a non-zero violation of the inequality $\mathcal{K}_{2 \mapsto 1}$. Therefore, optimization of $P(a_i,b_j|A_i,B_j)$ under the constraints $\mathcal{K}_{n \mapsto 1} = \mathcal{K}^{MR}_{n \mapsto 1}+ \epsilon$ enables one to determine the 
maximum value of $P(a_i,b_j|A_i,B_j)$ 
 for a certain amount of violation $\epsilon$. The other constraints $P(a_i,b_j|A_i,B_j) \geq 0$ and $\sum_{a_i,b_j} P(a_i,b_j|A_i,B_j) =1 ~~\forall A_i,B_j$ can be easily understood from the properties of a valid probability distribution. The last constraint that $P(a_i,b_j|A_i,B_j) ~\text{satisfies NSIT}$ implies that the probability distribution $P(a_i,b_j|A_i,B_j)$  satisfies the NSIT condition given by $P(b_j|B_j)=P(b_j|A_i,B_j)  ~\forall A_i,B_j,b_j$.

We are now interested to obtain a lower bound on the min-entropy $H_{\infty}(a_i,b_j|A_i,B_j)$ as a function of $\mathcal{K}_{n \mapsto 1}$, {\it i.e.}, we need to derive an inequality of the form $H_{\infty}(a_i,b_j|A_i,B_j) \geq f(\mathcal{K}_{n \mapsto 1})$. Below we provide a general lower bound of $H_{\infty}(a_i,b_j|A_i,B_j)$ from the perspective of no-signalling in time conditions. Using linear programming and imposing NSIT conditions, it can be shown that solving Eq.\eqref{abc} one can obtain $P^*(a_i,b_j|A_i,B_j) \leq \alpha \mathcal{K}_{n \mapsto 1} + \beta$, where $\alpha$ and $\beta$ in general may depend on $a_i,b_j,A_i,B_j$. 

\begin{figure}[t]
    \centering
    \includegraphics[scale=0.37]{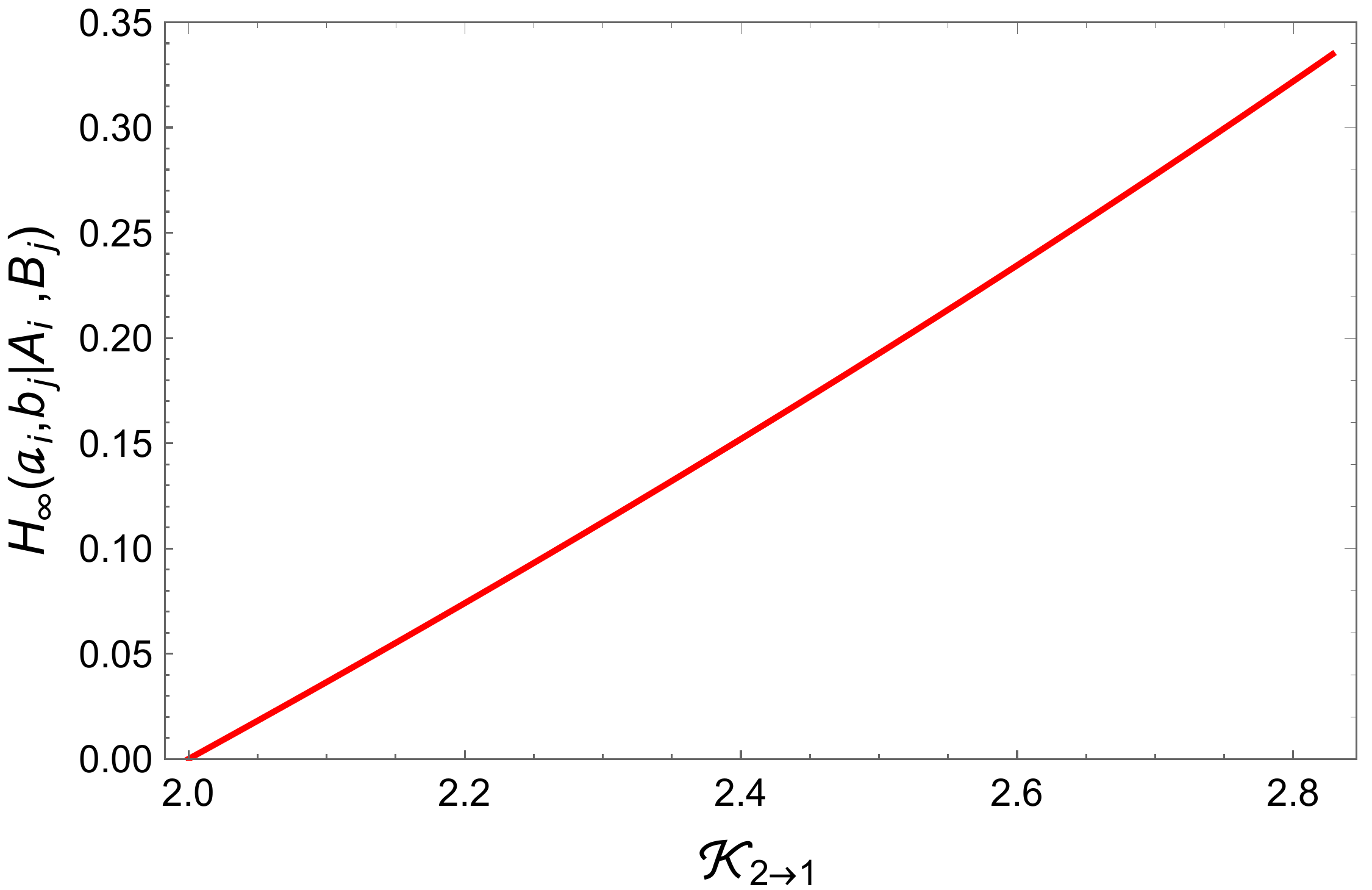}
    \caption{Min entropy $H_{\infty}(a_i,b_j|A_i,B_j)$ is plotted with $\mathcal{K}_{2 \mapsto 1}$}
    \label{fig1}
\end{figure}

For $2 \mapsto 1$ RAC, in the case of the classical strategy, a deterministic point can achieve $P(a_i,b_j|A_i,B_j)$ upto $1$, {\it i.e.}, $P^*(a_i,b_j|A_i,B_j) \leq 1$ when $\mathcal{K}_{2 \mapsto 1} = 2$, and for the no signaling (in time) box (which is equivalent to the Popescu-Rorlich box for spatial correlation), $P(a_i,b_j|A_i,B_j) =1/2$ {\it i.e.,} when $\mathcal{K}_{2 \mapsto 1} = 4$,  $P^*(a_i,b_j|A_i,B_j) \leq 1/2$ . Therefore, analyzing the above inequalities one can obtain the values of $\alpha$ and $\beta$ to be $-1/4$ and $3/2$, respectively \cite{randomness,rand'16}.
Hence,
\begin{align}
    P^*(a_i,b_j|A_i,B_j) \leq \frac{3}{2} - \frac{\mathcal{K}_{2 \mapsto 1}}{4} ~~\text{or} \nonumber \\
    H_{\infty}(a_i,b_j|A_i,B_j) \geq  -\log_2 [\frac{3}{2} - \frac{\mathcal{K}_{2 \mapsto 1}}{4}].
\end{align}
In fig.\eqref{fig1}, a graphical representation of the lower bound of $H_{\infty}(a_i,b_j|A_i,B_j)$ corresponding to $2 \mapsto 1$ RAC is provided.

\begin{figure}[t]
    \centering
    \includegraphics[scale=0.37]{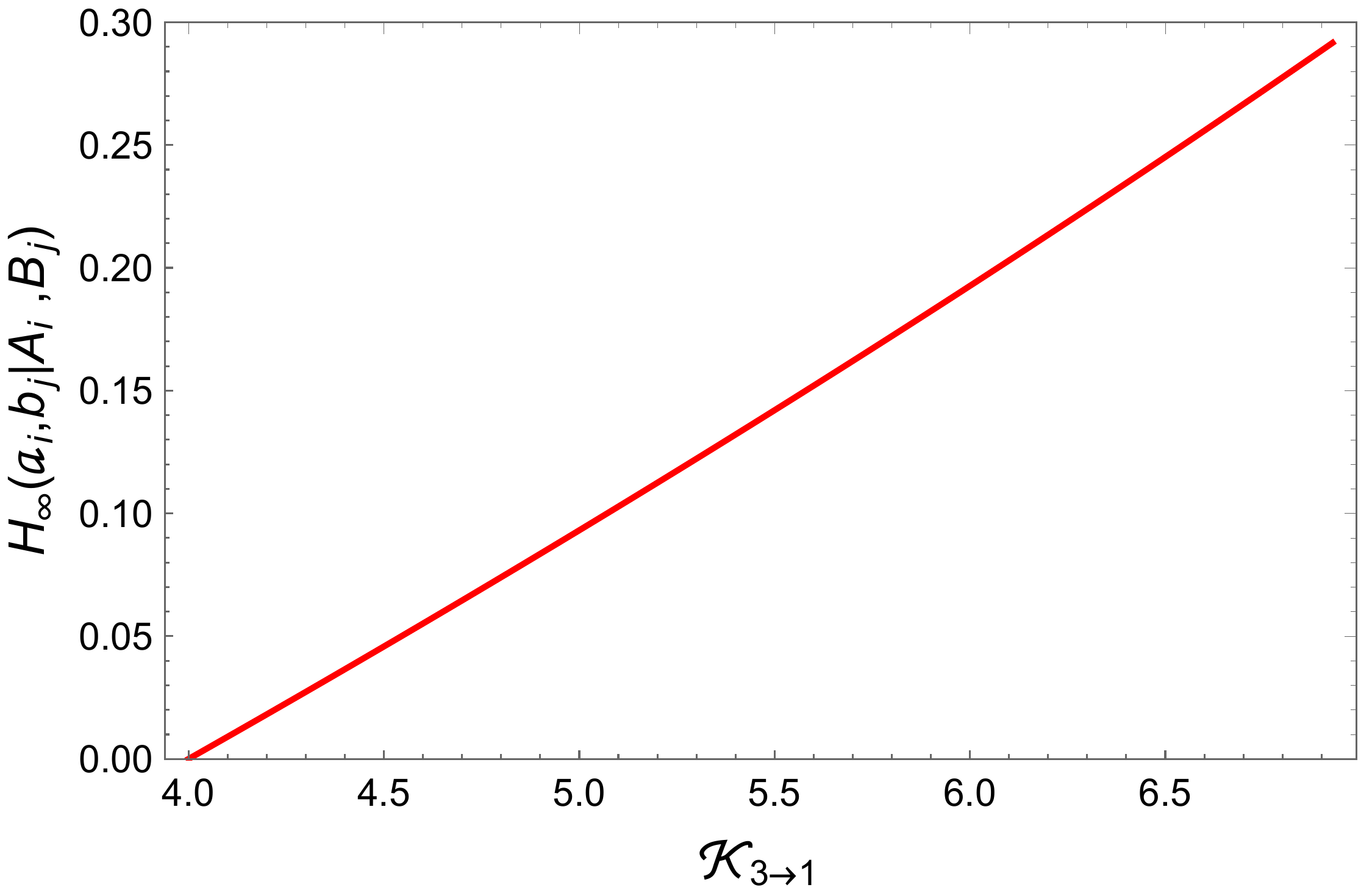}
    \caption{Min entropy $H_{\infty}(a_i,b_j|A_i,B_j)$ is plotted with $\mathcal{K}_{3 \mapsto 1}$}
    \label{fig2}
\end{figure}

\begin{figure}[t]
    \centering
    \includegraphics[scale=0.37]{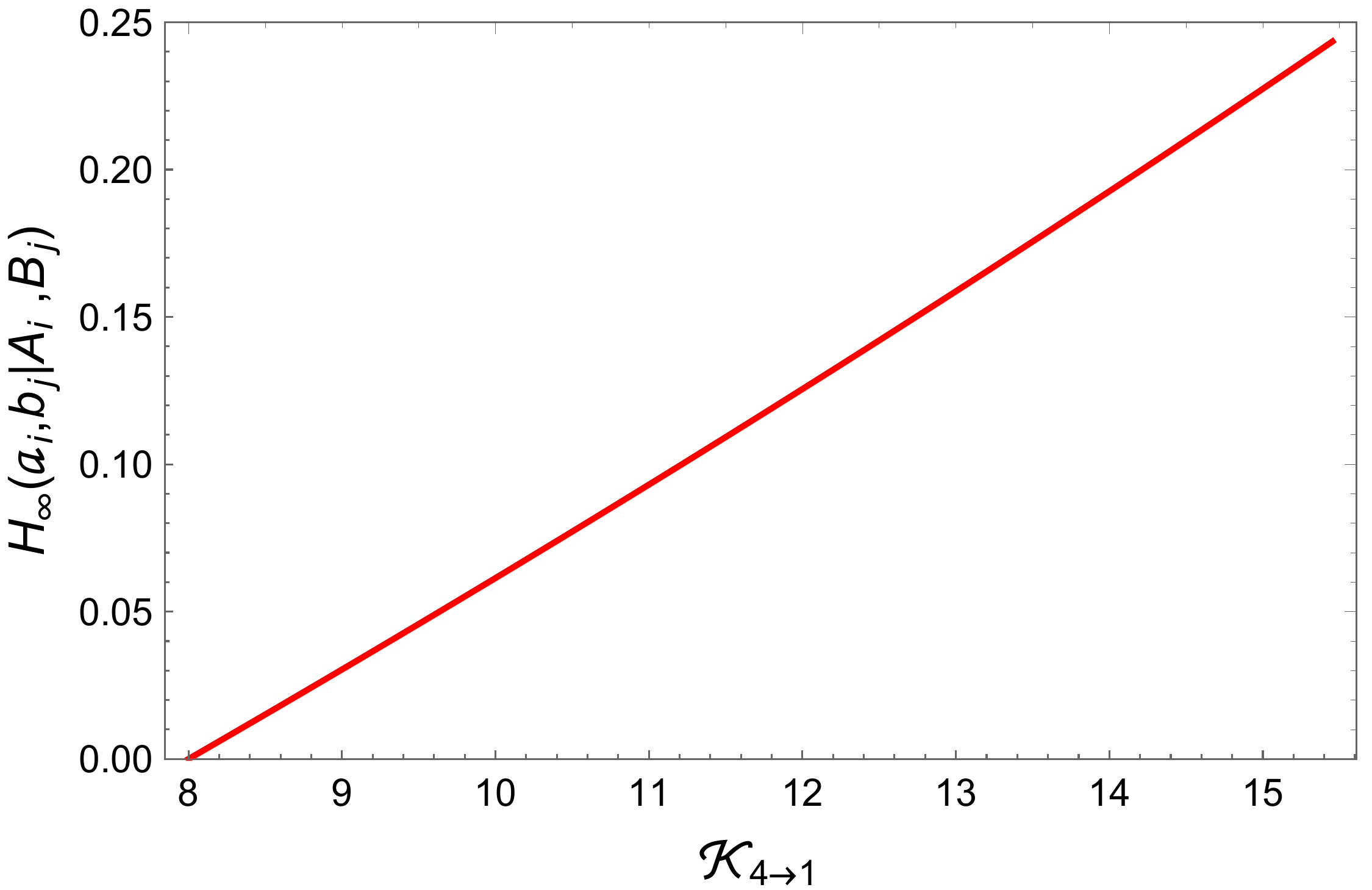}
    \caption{Min entropy $H_{\infty}(a_i,b_j|A_i,B_j)$ is plotted with $\mathcal{K}_{4 \mapsto 1}$}
    \label{fig3}
\end{figure}

A similar approach based on NSIT conditions can also be applied to other $n \mapsto 1$ RAC. For the case of $3 \mapsto 1$ RAC, the no-signalling polytope can achieve the value of $\mathcal{K}_{3 \mapsto 1}$ up to $12$ and for $4 \mapsto 1$ the value of  $\mathcal{K}_{4 \mapsto 1}$ up to $32$. Therefore, solving linear equations one can obtain $P^*(a_i,b_j|A_i,B_j) \leq \frac{5}{4} - \frac{\mathcal{K}_{3 \mapsto 1}}{16}$ and $P^*(a_i,b_j|A_i,B_j) \leq \frac{7}{6} - \frac{\mathcal{K}_{4 \mapsto 1}}{48}$ for $3 \mapsto 1$ and $4 \mapsto 1$ RAC, respectively. The lower bound of $H_{\infty}(a_i,b_j|A_i,B_j)$ corresponding to $\mathcal{K}_{3 \mapsto 1}$ and $\mathcal{K}_{4 \mapsto 1}$ RAC are plotted in fig.\eqref{fig2} and fig.\eqref{fig3} respectively. Here, it may be pertinent to mention that if instead of $P^*(a_i,b_j|A_i,B_j)$, randomness is certified from $P^*(b_j|A_i,B_j,a_i)$, then one  obtains the maximum value of $H_{\infty}(a_i,b_j|A_i,B_j)$ to be $1$ for any $n \mapsto 1$ RAC, because $P^*(b_j|A_i,B_j,a_i)$ is $1/2$ in this case irrespective of the value of $n$. Below, we summarize our key findings for the generation of genuine randomness based on our protocol.

$\bullet$ {\it \textbf{Result 2:-} } Any violation of the noninvasive-realist model of the temporal inequalities with initially prepared input state $\mathbb{I}/2$, or equivalently, any non-zero quantum advantage of $n \mapsto 1$ RAC with SR can be exploited to generate genuine randomness. It might be noted here that previous results to generate genuine randomness exploiting RAC do not guaranty genuine randomness for any non-zero quantum advantage of $n \mapsto 1$ RAC \cite{qkd,randomness1}.

\section{Conclusions} \label{s5}
Despite the fact that the random access code has enormous applications in information processing and communication tasks, the fundamental cause behind this advantage was hitherto not well understood. Here we showed that any non-zero quantum advantage of RAC with shared randomness necessarily violates a noninvasive-realist model. We proposed temporal inequalities corresponding to each $n \mapsto 1$ RAC with SR using the assumption of {\it realism} and {\it noninvasive measurability}. We have then established the fact that the maximum success probability of each $n \mapsto 1$ RAC with best classical strategy is connected to the maximum noninvasive-realist bound. 
Moreover, any non-zero quantum advantage of a $n \mapsto 1$ RAC was equivalent to the violation of the corresponding temporal inequality. 

Next, using an alternative derivation of the noninvasive-realist model, we showed that any non-zero advantage of RAC can be used to certify genuine randomness. This is particularly significant as all the previously proposed protocols based on RAC do not exhibit genuine randomness for the arbitrary quantum advantage of RAC \cite{rangen,qkd,randomness1}.
Before concluding a few remarks are in order. The maximum success probability using quantum strategy for a general $n \mapsto 1$ RAC is hard to compute for large $n$, and numerical strategies may be needed to tackle this problem. Finally, it may be reemphasized that our proposed protocol based on LGI violation is amenable for experimental realization \cite{urbasi1}, and hence, the generation of genuine randomness without entanglement based on our protocol might be exemplary for practical purposes. 

\section{Acknowledgements} SB and ASM acknowledge support of the Project No. 
DST/ICPS/QuEST/2018/98 of the Department of Science and Technology, Government of India. AGM acknowledges fruitful discussions with Debarshi Das and Bihalan Bhattacharya of SNBNCBS, Kolkata. SM acknowledges the support from the Ministry of Science and Technology, Taiwan (Grant No. MOST 111-2119-M-008-002).

\appendix

\section{Derivation of the classical bound for the temporal inequality using macrorealism} \label{appendixb}

\subsection{Ontic model of $2 \mapsto 1$ RAC:}\label{ap1}
The average of a quantum operator in the Heisenberg picture can be written as an average over a set of hidden variables $\lambda$. The role of the initial state $\rho(\lambda)$ is to provide a probability distribution on the set of hidden variables, called the ontic state. The average of an observable A (measurement carried out at time t can be written as
\begin{equation}
    \langle A_t\rangle = \int d\lambda A_t(\lambda)\rho(\lambda)
\end{equation}
where A$_t(\lambda$) is the value taken by the observable on the hidden variable $\lambda$. The correlation
between two observables $A_{t_i}$ (measured at some time $t_i$), $B_{t_j}$ (measured at some later time $t_j$) is given by
\begin{equation}
\label{aa2}
    \langle A_{t_i}B_{t_j}\rangle = \int d\lambda A_{t_i}(\lambda)B_{t_j}(\lambda)\rho(\lambda\mid A_{t_i})
\end{equation}
In general, we can always ignore the effect of final measurement due to noninvasive
measurability (NIM). NIM can be defined as $\rho(\lambda\mid A_{t_i},B_{t_j},\dots)$ = $\rho(\lambda)$, i.e., a measurement does not change the distribution of $\lambda$. In Eq.\eqref{aa2}, $\rho(\lambda)$ not depends on observable $B_{t_j}$ due to observable $B_{t_j}$ being measured after the measurement of observable $A_{t_i}$.

Let us take Alice's preparations to be eigenstates of observables  $\{A_1, A_2\}$ and Bob's measurements to be $\{B_1, B_2\}$. Let, measurement A be carried out at some time $t_i$, and measurement B be carried out at some later time $t_{j}$. Now, imposing the conditions of realism and NIM, we obtain
\begin{align}\label{A(1)B(1)}
&\langle A_1B_1\rangle +\langle A_2B_1\rangle\nonumber\\
&= \int d\lambda [A_1(\lambda)B_1(\lambda)\rho(\lambda\mid A_1)+A_2(\lambda)B_1(\lambda)\rho(\lambda\mid A_2)]\nonumber\\
&= \int d\lambda A_1(\lambda)B_1(\lambda)[1\mp A_2(\lambda)B_2(\lambda)]\rho(\lambda\mid A_1)\nonumber\\
&+\int d\lambda A_2(\lambda)B_1(\lambda)[1\pm A_1(\lambda)B_2(\lambda)]\rho(\lambda\mid A_2).
\end{align}
Taking the modulus on both sides and using the triangle inequality we obtain,
\begin{align}
&|\langle A_1B_1\rangle +\langle A_2B_1\rangle|\nonumber\\
&\le 2\pm [ \int d\lambda A_1(\lambda)B_2(\lambda)\rho(\lambda\mid A_1)\nonumber\\
&- \int d\lambda A_2(\lambda)B_2(\lambda)\rho(\lambda\mid A_2)].
\end{align}
Now invoking NIM, we have,
\begin{align}
&|\langle A_1B_1\rangle +\langle A_2B_1\rangle|\nonumber\\
&\mp [\langle A_1B_2\rangle -\langle A_2B_2\rangle]\le 2.\nonumber\\
\text{or},\nonumber\\
&\mathcal{K}_{2 \mapsto 1} \le 2.
\end{align}
This is the four term Leggett-Garg inequality.

\subsection{Ontic model of $3 \mapsto 1$ RAC :}\label{ap2}
Let us take Alice's preparations to be eigenstates of $\{A_1, A_2, A_3, A_4\}$ and Bob's measurements to be $\{B_1, B_2, B_3\}$. Now, following similar steps as in the derivation of the Bell inequality, we obtain the sum of four correlations,
\begin{align}\label{A(4)B(3)}
&\langle A_1B_1\rangle +\langle A_4B_1\rangle + \langle A_2B_1\rangle +\langle A_3B_1\rangle\nonumber\\
&= \int d\lambda [A_1(\lambda)B_1(\lambda)\rho(\lambda\mid A_1)+A_4(\lambda)B_1(\lambda)\rho(\lambda\mid A_4) +\nonumber\\
&A_2(\lambda)B_1(\lambda)\rho(\lambda\mid A_2)+
A_3(\lambda)B_1(\lambda)\rho(\lambda\mid A_3)]\nonumber\\
&= \int d\lambda A_1(\lambda)B_1(\lambda)[1\mp A_4(\lambda)(B_2(\lambda)+B_3(\lambda))]\rho(\lambda\mid A_1)\nonumber\\
&+\int d\lambda A_4(\lambda)B_1(\lambda)[1\pm A_1(\lambda)(B_2(\lambda)+B_3(\lambda))]\rho(\lambda\mid A_4)\nonumber\\
&+ \int d\lambda A_2(\lambda)B_1(\lambda)[1\mp A_3(\lambda)(B_2(\lambda)-B_3(\lambda))]\rho(\lambda\mid A_2)\nonumber\\
&+\int d\lambda A_3(\lambda)B_1(\lambda)[1\pm A_2(\lambda)(B_2(\lambda)-B_3(\lambda))]\rho(\lambda\mid A_3)
\end{align}
Now, we take the modulus of
both sides and use the triangle inequality to obtain,
\begin{align}\label{A(4)B(2)}
&|\langle A_1B_1\rangle +\langle A_4B_1\rangle + \langle A_2B_1\rangle +\langle A_3B_1\rangle\nonumber|\\
&\le 4\pm [ \int d\lambda A_1(\lambda)(B_2(\lambda)+B_3(\lambda))\rho(\lambda\mid A_1)\nonumber\\
&- \int d\lambda A_4(\lambda)(B_2(\lambda)+B_3(\lambda))\rho(\lambda\mid A_4)\nonumber\\
&+ \int d\lambda A_2(\lambda)(B_2(\lambda)-B_3(\lambda))\rho(\lambda\mid A_2)\nonumber\\
&- \int d\lambda A_3(\lambda)(B_2(\lambda)-B_3(\lambda))\rho(\lambda\mid A_3)].
\end{align}
Invoking NIM, we have,
\begin{align}\label{A(4)B(1)}
&|\langle A_1B_1\rangle +\langle A_4B_1\rangle + \langle A_2B_1\rangle +\langle A_3B_1\rangle|\nonumber\\
&\mp [\langle A_1(B_2 + B_3)\rangle -\langle A_4(B_2 + B_3)\rangle + \langle A_2(B_2 - B_3)\rangle\nonumber\\
&-\langle A_3(B_2 - B_3)]\le 4.\nonumber\\
\text{or},\nonumber\\
&\mathcal{K}_{3 \mapsto 1} \le 4.
\end{align}

\subsection{Ontic model of $4 \mapsto 1$ RAC :} \label{ap3}
Similarly, for $4 \mapsto 1$ RAC, here Alice has 8 preparations to be eigenstates of $\{A_1, A_2,..., A_8\}$ and Bob has 4 measurements to be $\{B_1, B_2, B_3, B_4\}$. Now using a similar procedure presented for deriving the classical bound corresponding to $2 \mapsto 1$ RAC, we can obtain
\begin{align}\label{A(8)B(4)}
&\langle A_1B_1\rangle +\langle A_5B_1\rangle + \langle A_1B_3\rangle +\langle A_2B_3\rangle + \langle A_2B_1\rangle +\langle A_6B_1\rangle\nonumber\\ 
&+ \langle A_3B_1\rangle +\langle A_7B_1\rangle + \langle A_4B_1\rangle +\langle A_8B_1\rangle +\langle A_5B_3\rangle \nonumber\\
&+\langle A_6B_3\rangle - \langle A_4B_3\rangle -\langle A_7B_3\rangle - \langle A_8B_3\rangle -\langle A_3B_3\rangle\nonumber\\
&= \int d\lambda [A_1(\lambda)B_1(\lambda)\rho(\lambda\mid A_1)+A_5(\lambda)B_1(\lambda)\rho(\lambda\mid A_5) +\nonumber\\
&A_1(\lambda)B_3(\lambda)\rho(\lambda\mid A_1)+
A_2(\lambda)B_3(\lambda)\rho(\lambda\mid A_2) +\nonumber\\
&A_2(\lambda)B_1(\lambda)\rho(\lambda\mid A_2)+A_6(\lambda)B_1(\lambda)\rho(\lambda\mid A_6) +\nonumber\\
&A_3(\lambda)B_1(\lambda)\rho(\lambda\mid A_3)
+A_7(\lambda)B_1(\lambda)\rho(\lambda\mid A_7) +\nonumber\\
&A_4(\lambda)B_1(\lambda)\rho(\lambda\mid A_4)+A_8(\lambda)B_1(\lambda)\rho(\lambda\mid A_8) +\nonumber\\
&A_5(\lambda)B_3(\lambda)\rho(\lambda\mid A_5)
+A_6(\lambda)B_3(\lambda)\rho(\lambda\mid A_6) -\nonumber\\
&A_4(\lambda)B_3(\lambda)\rho(\lambda\mid A_4)-A_7(\lambda)B_3(\lambda)\rho(\lambda\mid A_7) -\nonumber\\
&A_8(\lambda)B_3(\lambda)\rho(\lambda\mid A_8)
-A_3(\lambda)B_3(\lambda)]\rho(\lambda\mid A_3)]\nonumber\\
&= \int d\lambda A_1(\lambda)B_1(\lambda)[1\mp A_5(\lambda)B_2(\lambda)]\rho(\lambda\mid A_1)\nonumber\\
&+\int d\lambda A_5(\lambda)B_1(\lambda)[1\pm A_1(\lambda)B_2(\lambda)]\rho(\lambda\mid A_5)\nonumber\\
&+ \int d\lambda A_1(\lambda)B_3(\lambda)[1\mp A_2(\lambda)B_4(\lambda)]\rho(\lambda\mid A_1)\nonumber\\
&+\int d\lambda A_2(\lambda)B_3(\lambda)[1\pm A_1(\lambda)B_4(\lambda)]\rho(\lambda\mid A_2)\nonumber\\
&+ \int d\lambda A_2(\lambda)B_1(\lambda)[1\mp A_6(\lambda)B_2(\lambda)]\rho(\lambda\mid A_2)\nonumber\\
&+\int d\lambda A_6(\lambda)B_1(\lambda)[1\pm A_2(\lambda)B_2(\lambda)]\rho(\lambda\mid A_6)\nonumber\\
&+ \int d\lambda A_3(\lambda)B_1(\lambda)[1\mp A_7(\lambda)B_2(\lambda)]\rho(\lambda\mid A_3)\nonumber\\
&+\int d\lambda A_7(\lambda)B_1(\lambda)[1\pm A_3(\lambda)B_2(\lambda)]\rho(\lambda\mid A_7)\nonumber\\
&+ \int d\lambda A_4(\lambda)B_1(\lambda)[1\mp A_8(\lambda)B_2(\lambda)]\rho(\lambda\mid A_4)\nonumber\\
&+\int d\lambda A_8(\lambda)B_1(\lambda)[1\pm A_4(\lambda)B_2(\lambda)]\rho(\lambda\mid A_8)\nonumber\\
&+ \int d\lambda A_5(\lambda)B_3(\lambda)[1\mp A_6(\lambda)B_4(\lambda)]\rho(\lambda\mid A_5)\nonumber\\
&+\int d\lambda A_6(\lambda)B_3(\lambda)[1\pm A_5(\lambda)B_4(\lambda)]\rho(\lambda\mid A_6)\nonumber\\
&- \int d\lambda A_4(\lambda)B_3(\lambda)[1\mp A_7(\lambda)B_4(\lambda)]\rho(\lambda\mid A_4)\nonumber\\
&-\int d\lambda A_7(\lambda)B_3(\lambda)[1\pm A_4(\lambda)B_4(\lambda)]\rho(\lambda\mid A_7)\nonumber\\
&- \int d\lambda A_8(\lambda)B_3(\lambda)[1\mp A_3(\lambda)B_4(\lambda)]\rho(\lambda\mid A_8)\nonumber\\
&-\int d\lambda A_3(\lambda)B_3(\lambda)[1\pm A_8(\lambda)B_4(\lambda)]\rho(\lambda\mid A_3)
\end{align}
Taking the modulus on both sides and using the triangle inequality we obtain,
\begin{align}
&|\langle A_1B_1\rangle +\langle A_5B_1\rangle + \langle A_1B_3\rangle +\langle A_2B_3\rangle + \langle A_2B_1\rangle +\nonumber\\
&\langle A_6B_1\rangle + \langle A_3B_1\rangle +\langle A_7B_1\rangle + \langle A_4B_1\rangle +\langle A_8B_1\rangle +\nonumber\\
&\langle A_5B_3\rangle +\langle A_6B_3\rangle - \langle A_4B_3\rangle -\langle A_7B_3\rangle - \langle A_8B_3\rangle\nonumber\\
&-\langle A_3B_3\rangle|\le 8\pm [ \int d\lambda A_1(\lambda)B_2(\lambda)\rho(\lambda\mid A_1)\nonumber\\
&- \int d\lambda A_5(\lambda)B_2(\lambda)\rho(\lambda\mid A_5) + \int d\lambda A_1(\lambda)B_4(\lambda)\rho(\lambda\mid A_1)\nonumber\\
&- \int d\lambda A_2(\lambda)B_4(\lambda)\rho(\lambda\mid A_2) + \int d\lambda A_2(\lambda)B_2(\lambda)\rho(\lambda\mid A_2)\nonumber\\
&- \int d\lambda A_6(\lambda)B_2(\lambda)\rho(\lambda\mid A_6) + \int d\lambda A_3(\lambda)B_2(\lambda)\rho(\lambda\mid A_3)\nonumber\\
&- \int d\lambda A_7(\lambda)B_2(\lambda)\rho(\lambda\mid A_7) + \int d\lambda A_4(\lambda)B_2(\lambda)\rho(\lambda\mid A_4)\nonumber\\
&- \int d\lambda A_8(\lambda)B_2(\lambda)\rho(\lambda\mid A_8) + \int d\lambda A_5(\lambda)B_4(\lambda)\rho(\lambda\mid A_5)\nonumber\\
&- \int d\lambda A_6(\lambda)B_4(\lambda)\rho(\lambda\mid A_6) + \int d\lambda A_7(\lambda)B_4(\lambda)\rho(\lambda\mid A_4)\nonumber\\
&- \int d\lambda A_4(\lambda)B_4(\lambda)\rho(\lambda\mid A_7) + \int d\lambda A_3(\lambda)B_4(\lambda)\rho(\lambda\mid A_8)\nonumber\\
&- \int d\lambda A_8(\lambda)B_4(\lambda)\rho(\lambda\mid A_3)].
\end{align}
Invoking NIM, we have,
\begin{align}
&|\langle A_1B_1\rangle +\langle A_5B_1\rangle + \langle A_1B_3\rangle +\langle A_2B_3\rangle + \langle A_2B_1\rangle +\nonumber\\
&\langle A_6B_1\rangle + \langle A_3B_1\rangle +\langle A_7B_1\rangle + \langle A_4B_1\rangle +\langle A_8B_1\rangle +\nonumber\\
&\langle A_5B_3\rangle +\langle A_6B_3\rangle - \langle A_4B_3\rangle -\langle A_7B_3\rangle - \langle A_8B_3\rangle\nonumber\\
&-\langle A_3B_3\rangle|\mp [\langle A_1B_2\rangle -\langle A_5B_2\rangle + \langle A_1B_4\rangle -\langle A_2B_4\rangle\nonumber\\
&+ \langle A_2B_2\rangle - \langle A_6B_2\rangle + \langle A_3B_2\rangle -\langle A_7B_2\rangle + \langle A_4B_2\rangle-\nonumber\\
&\langle A_8B_2\rangle +\langle A_5B_4\rangle -\langle A_6B_4\rangle - \langle A_4B_4\rangle +\langle A_7B_4\rangle -\nonumber\\
&\langle A_8B_4\rangle +\langle A_3B_4\rangle]\le 8.\nonumber\\
\text{or},\nonumber\\
&\mathcal{K}_{4 \mapsto 1} \le 8.
\end{align}

\subsection{Ontic model of $n \mapsto 1$ RAC :} \label{apn}
Here Alice has $2^{n-1}$ preparations which are the eigenstates corresponding to $\{A_1, A_2,..., A_{2^{n-1}}\}$ and Bob has n measurements, $\{B_1, B_2, B_3,..., B_n\}$. Following Eq.$\eqref{kn}$ the term $\mathcal{K}_{n \mapsto 1}$ can be explicitly written as,
\begin{align}
&\mathcal{K}_{n \mapsto 1} =\langle A_1B_1\rangle +\langle A_1B_2\rangle +\langle A_1B_3\rangle +\langle A_1B_4\rangle +\dots +\nonumber\\
&\langle A_1B_n\rangle +\langle A_2B_1\rangle 
+\langle A_2B_2\rangle +\langle A_2B_3\rangle +\langle A_2B_4\rangle -\langle A_2B_n\rangle\nonumber\\
&+\langle A_3B_1\rangle+\langle A_3B_2\rangle+\langle A_3B_3\rangle+\langle A_3B_4\rangle+\langle A_3B_n\rangle\nonumber\\
& +\dots+ \langle A_{2^{n-1}-1}B_1\rangle -\langle A_{2^{n-1}-1}B_2\rangle-\langle A_{2^{n-1}-1}B_3\rangle\nonumber\\
&-\dots+\langle A_{2^{n-1}-1}B_n\rangle+\langle A_{2^{n-1}}B_1\rangle -\langle A_{2^{n-1}}B_2\rangle\nonumber\\
&-\langle A_{2^{n-1}}B_3\rangle-\dots-
\langle A_{2^{n-1}}B_{n-3}\rangle-\langle A_{2^{n-1}}B_{n-1}\rangle\nonumber\\
&-\langle A_{2^{n-1}}B_n\rangle .
\end{align}

Now, we can obtain the  classical bound for $n \mapsto 1$ RAC if we adopt a similar procedure as presented for deriving the classical bound corresponding to $2 \mapsto 1$ RAC. Let us now derive the classical bound explicitly when $n$ is even.
\begin{align}\label{A(8)B(n)}
&\langle A_1B_1\rangle +\langle A_1B_3\rangle +\dots +\langle A_2B_1\rangle +\langle A_2B_3\rangle +\dots +\nonumber\\
&\langle A_3B_1\rangle +\langle A_3B_3\rangle 
+\dots + \langle A_{2^{n-1}-1}B_1\rangle -\nonumber\\
&\langle A_{2^{n-1}-1}B_3\rangle -\dots + \langle A_{2^{n-1}}B_1\rangle -\langle A_{2^{n-1}}B_3\rangle -\dots\nonumber\\
&-\langle A_{2^{n-1}}B_{n-3}\rangle-\langle A_{2^{n-1}}B_{n-1}\rangle\nonumber\\
&= \int d\lambda [A_1(\lambda)B_1(\lambda)\rho(\lambda\mid A_1)+A_1(\lambda)B_3(\lambda)\rho(\lambda\mid A_1) \nonumber\\
&+\dots+ A_2(\lambda)B_1(\lambda)\rho(\lambda\mid A_2)+
A_2(\lambda)B_3(\lambda)\rho(\lambda\mid A_2)+\nonumber\\ 
&\dots+A_3(\lambda)B_1(\lambda)\rho(\lambda\mid A_3)+A_3(\lambda)B_3(\lambda)\rho(\lambda\mid A_3)\nonumber\\ 
&+\dots+
A_{2^{n-1}-1}(\lambda)B_1(\lambda)\rho(\lambda\mid A_{2^{n-1}-1})\nonumber\\
&-A_{2^{n-1}-1}(\lambda)B_3(\lambda)\rho(\lambda\mid A_{2^{n-1}-1}) -\dots+\nonumber\\
&A_{2^{n-1}}(\lambda)B_1(\lambda)\rho(\lambda\mid A_{2^{n-1}})-A_{2^{n-1}}(\lambda)B_3(\lambda)\rho(\lambda\mid A_{2^{n-1}})\nonumber\\
&-\dots-A_{2^{n-1}}(\lambda)B_{n-3}(\lambda)\rho(\lambda\mid A_{2^{n-1}})\nonumber\\
&-A_{2^{n-1}}(\lambda)B_{n-1}(\lambda)\rho(\lambda\mid A_{2^{n-1}})]\nonumber\\
&= \int d\lambda A_1(\lambda)B_1(\lambda)[1\mp A_{2^{n-1}}(\lambda)B_2(\lambda)]\rho(\lambda\mid A_{2^{n-1}})\nonumber\\
&+\int d\lambda A_2(\lambda)B_1(\lambda)[1\mp A_{2^{n-1}-1}(\lambda)B_2(\lambda)]\rho(\lambda\mid A_{2^{n-1}-1})\nonumber\\
&+ \int d\lambda A_1(\lambda)B_{n-1}(\lambda)[1\mp A_2(\lambda)B_n(\lambda)]\rho(\lambda\mid A_1)\nonumber\\
&+\int d\lambda A_2(\lambda)B_{n-1}(\lambda)[1\pm A_1(\lambda)B_n(\lambda)]\rho(\lambda\mid A_2)\nonumber\\
&+\dots+ \int d\lambda A_{2^{n-1}-1}(\lambda)B_1(\lambda)[1\pm\nonumber\\ &A_2(\lambda)B_2(\lambda)]\rho(\lambda\mid A_{2^{n-1}-1})\nonumber\\
&+\int d\lambda A_{2^{n-1}}(\lambda)B_1(\lambda)[1\pm\nonumber\\ &A_1(\lambda)B_2(\lambda)]\rho(\lambda\mid A_{2^{n-1}})\nonumber\\
&+\dots- \int d\lambda A_{2^{n-1}-1}(\lambda)B_{n-1}(\lambda)[1\mp\nonumber\\ &A_{2^{n-1}}(\lambda)B_n(\lambda)]\rho(\lambda\mid A_{2^{n-1}-1})\nonumber\\
&-\int d\lambda A_{2^{n-1}}(\lambda)B_{n-1}(\lambda)[1\pm\nonumber\\
&A_{2^{n-1}-1}(\lambda)B_n(\lambda)]\rho(\lambda\mid A_{2^{n-1}})\nonumber\\
\end{align}
Taking the modulus on both sides and using the triangle inequality we obtain,
\begin{align}
&|\langle A_1B_1\rangle +\langle A_1B_3\rangle +\dots +\langle A_2B_1\rangle +\langle A_2B_3\rangle +\dots +\nonumber\\
&\langle A_3B_1\rangle +\langle A_3B_3\rangle 
+\dots + \langle A_{2^{n-1}-1}B_1\rangle -\nonumber\\
&\langle A_{2^{n-1}-1}B_3\rangle -\dots + \langle A_{2^{n-1}}B_1\rangle -\langle A_{2^{n-1}}B_3\rangle\nonumber\\
&-\dots-\langle A_{2^{n-1}}B_{n-3}\rangle-\langle A_{2^{n-1}}B_{n-1}\rangle|\le 2^{n-1}\nonumber\\
&\pm [ \int d\lambda A_1(\lambda)B_2(\lambda)\rho(\lambda\mid A_1)\nonumber\\
&+ \int d\lambda A_2(\lambda)B_2(\lambda)\rho(\lambda\mid A_2)+\dots\nonumber\\
&+ \int d\lambda A_1(\lambda)B_n(\lambda)\rho(\lambda\mid A_1)\nonumber\\
&- \int d\lambda A_2(\lambda)B_n(\lambda)\rho(\lambda\mid A_2)+\dots\nonumber\\ 
&- \int d\lambda A_{2^{n-1}-1}(\lambda)B_2(\lambda)\rho(\lambda\mid A_{2^{n-1}-1})\nonumber\\
&- \int d\lambda A_{2^{n-1}}(\lambda)B_2(\lambda)\rho(\lambda\mid A_{2^{n-1}})-\dots\nonumber\\ 
&- \int d\lambda A_{2^{n-1}-1}(\lambda)B_n(\lambda)\rho(\lambda\mid A_{2^{n-1}-1})\nonumber\\
&- \int d\lambda A_{2^{n-1}}(\lambda)B_n(\lambda)\rho(\lambda\mid A_{2^{n-1}})].
\end{align}
Invoking NIM, we have,
\begin{align}
&|\langle A_1B_1\rangle +\langle A_1B_3\rangle +\dots +\langle A_2B_1\rangle +\langle A_2B_3\rangle +\dots +\nonumber\\
&\langle A_3B_1\rangle +\langle A_3B_3\rangle 
+\dots + \langle A_{2^{n-1}-1}B_1\rangle -\nonumber\\
&\langle A_{2^{n-1}-1}B_3\rangle -\dots + \langle A_{2^{n-1}}B_1\rangle -\langle A_{2^{n-1}}B_3\rangle\nonumber\\
&-\langle A_{2^{n-1}}B_{n-3}\rangle-\langle A_{2^{n-1}}B_{n-1}\rangle|\mp [\langle A_1B_2\rangle+\nonumber\\
&\langle A_1B_4\rangle +\dots+ \langle A_1B_n\rangle -\langle A_2B_2\rangle\nonumber\\
&+ \langle A_2B_4\rangle+\dots - \langle A_2B_n\rangle + \langle A_3B_2\rangle +\langle A_3B_4\rangle\nonumber\\ 
&+\dots+ \langle A_3B_n\rangle+\dots-
\langle A_{2^{n-1}-1}B_2\rangle -\dots+\nonumber\\
&\langle A_{2^{n-1}-1}B_n\rangle -\langle A_{2^{n-1}}B_2\rangle -\dots- \langle A_{2^{n-1}}B_n\rangle]\le 2^{n-1}.\nonumber\\
\text{or},\nonumber\\
&\mathcal{K}_{n \mapsto 1} \le 2^{n-1}.
\end{align}
Similarly, one can also obtain the same classical bound for $\mathcal{K}_{n \mapsto 1}$ when $n$ is odd.

\end{document}